\documentclass[journal,10pt]{IEEEtran}
\usepackage{cite}
\usepackage{tikz}
\usepackage{amsmath,amsfonts}
\usepackage{array} 
\usepackage{graphicx} 
\usepackage{enumitem} 
\usepackage{enumitem}
\usepackage{tabu,multirow}
\usepackage{algorithmic}
\usepackage{arydshln}  
\usepackage{tabularx}
\usepackage{array}
\usepackage[caption=false,font=normalsize,labelfont=sf,textfont=sf]{subfig}
\usepackage{textcomp}
\usepackage{stfloats}
\usepackage{url}
\usepackage{verbatim}
\usepackage{comment}
\usepackage{booktabs}  
\usepackage{array}
\newcolumntype{?}{!{\vrule width 1pt}}
\usepackage{graphicx}
\def\BibTeX{{\rm B\kern-.05em{\sc i\kern-.025em b}\kern-.08em
    T\kern-.1667em\lower.7ex\hbox{E}\kern-.125emX}}
\usepackage{balance}

\begin{document}

\title{Deep Learning in Wireless Communication Receiver: A Survey}

\author{Shadman~Rahman~Doha, Ahmed~Abdelhadi
\thanks{Shadman Rahman Doha is with the School of Electrical Engineering \& Computer Science University of North Dakota, Grand Forks, ND 58202, USA.}
\thanks{Ahmed Abdelhadi  is with the School of Electrical Engineering \& Computer Science University of North Dakota, Grand Forks, ND 58202, USA.}
}

\maketitle

\begin{abstract}
The design of wireless communication receivers to enhance signal processing in complex and dynamic environments is going through a transformation by leveraging deep neural networks (DNNs). Traditional wireless receivers depend on mathematical models and algorithms, which do not have the ability to adapt or learn from data. In contrast, deep learning-based receivers are more suitable for modern wireless communication systems because they can learn from data and adapt accordingly. This survey explores various deep learning architectures such as multilayer perceptrons (MLPs), convolutional neural networks (CNNs), recurrent neural networks (RNNs), generative adversarial networks (GANs), and autoencoders, focusing on their application in the design of wireless receivers. Key modules of a receiver such as synchronization, channel estimation, equalization, space-time decoding, demodulation, decoding, interference cancellation, and modulation classification are discussed in the context of advanced wireless technologies like orthogonal frequency division multiplexing (OFDM), multiple input multiple output (MIMO), semantic communication, task-oriented communication, and next-generation (Next-G) networks. The survey not only emphasizes the potential of deep learning-based receivers in future wireless communication but also investigates different challenges of deep learning-based receivers, such as data availability, security and privacy concerns, model interpretability, computational complexity, and integration with legacy systems.
\end{abstract}

\section{Introduction}
\label{sec:Intro}
Deep learning (DL) is emerging as a new paradigm for the design of wireless receivers, using DNNs to enhance performance in complex communication scenarios. Traditional wireless receivers are designed with static mathematical models and algorithms, which limit their effectiveness in unknown and highly dynamic channel conditions. On the other hand, DNN-based receivers can adaptively learn from the data, making it suitable for designing wireless receivers that are more efficient in modern wireless systems \cite{DeepReciverdataAug}, \cite{DeepReciverdatahybrid}. In this survey, we explore the diverse deep-learning approaches transforming the receiver design and overall signal processing. This survey serves as a valuable resource for researchers and industry professionals seeking to leverage DL in wireless communication receiver design. By providing an in-depth analysis of existing approaches, challenges, and emerging trends, this survey paves the way for future research and innovation in the field.

Various survey literature has studied the effectiveness of Machine Learning (ML) and DL in the wireless communication domain. An overview of the historical evolution of machine learning, from shallow to deep learning, is presented by the authors in \cite{DL_Overview}. A comprehensive analysis of DL applications in mobile and wireless networks is presented in \cite{DLforWireless}, where the authors focus on challenges like adversarial attacks, computational demands, and interpretability. The capability of DL in handling heterogeneous data and automatic feature extraction in diverse domains such as mobility, localization, network control, and security is also highlighted in this survey. Similarly, authors in \cite{8666641} emphasize the potential of DL in wireless communications by describing its role in designing end-to-end communication systems, spectrum awareness, wireless security, and enhancing system performance through techniques like autoencoders, GANs, and adversarial learning. In \cite{Aldossari2019MachineLF}, the authors underscore the utility of ML, especially DL, in tackling the complexities of 5G and internet of things (IoT) systems. The work categorizes ML approaches into supervised, unsupervised, and reinforcement learning for wireless channel modeling and estimation. Further, the advancements of DL in intelligent communication by reviewing its applications in cognitive radio (CR), edge computing (EC), channel measurement (CM), end-to-end encoder/decoder (EED), and visible light communication (VLC) are illustrated in \cite{9665363}.

Together, these studies illustrate the growing influence of DL and ML in optimizing wireless networks, advancing communication technologies, and addressing future challenges. However, a survey on different DL models used in wireless receivers highlighting the challenges addressed by the DL networks is not present. We have investigated different types of DNN architectures, such as MLPs, CNNs, RNNs, GANs, and autoencoders, to determine their unique abilities in addressing the challenges of traditional wireless communication. Our survey further investigates the implementation of DL in designing various modules of wireless receivers, including synchronization, channel estimation, equalization, space-time decoding, demodulation, channel decoding, decryption, and source decoding. We analyze how DL enhances traditional receiver functionalities and enables novel receiver designs that outperform conventional methods in various wireless communication scenarios. Moreover, application of DL-based receiver in different wireless communication systems such as semantic communication, task-oriented communication, OFDM, MIMO, and Next-G networks are explored in this study. We focus on highlighting the advantages and challenges of DL-based wireless receivers in these systems.

The rest of the paper is organized in the following sections. In Section \ref{sec: System Model}, the structure of a traditional wireless receiver is introduced. In this section, we discussed different wireless receiver modules and the methods used to optimize these individual components. Section \ref{sec:Deep Neural Networks Architectures} describes different DNN architectures and their application in the wireless communication domain. Section \ref{Sec: Application of Deep Learning based receivers} is a survey on the applications of deep-learning-based receivers in different wireless communication technologies. The challenges of DL implementation in wireless receivers are discussed in Section \ref{sec:Challenes}. Conclusions are presented in Section \ref{sec:conclusion}.

\section{Wireless Receiver System Model}
\label{sec: System Model}

\begin{figure*}[htbp]
        \centering
        \captionsetup{justification=centering}
        \includegraphics[width=.9\linewidth]{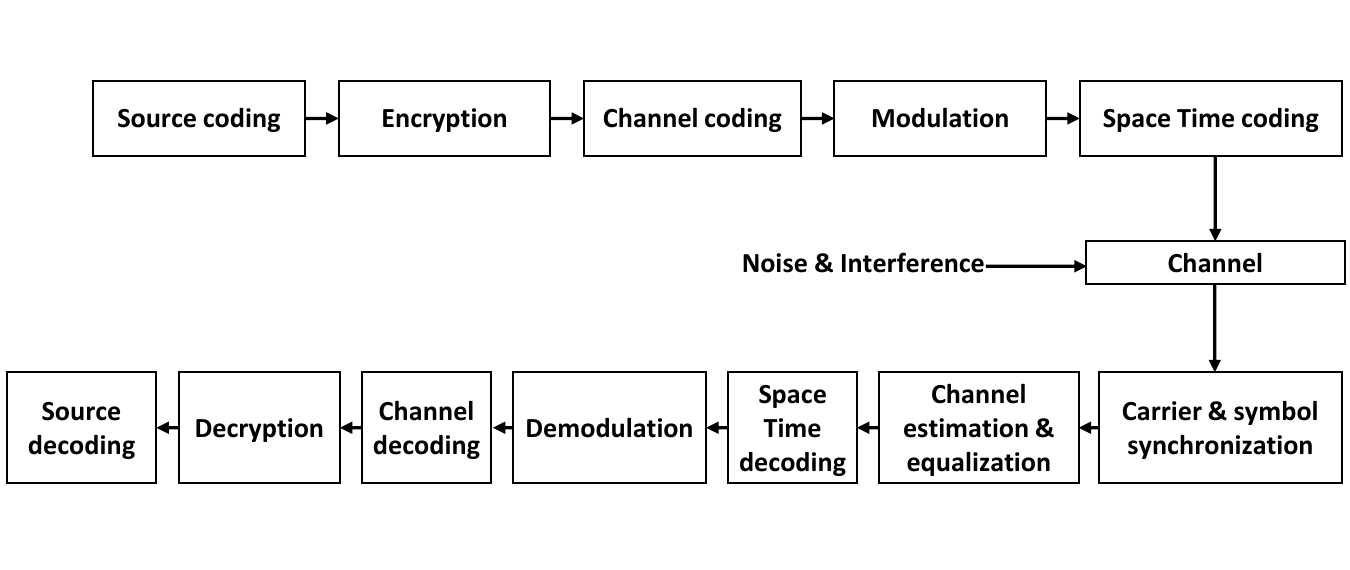}
        \centering
        \caption{System Model}
        
        \label{fig:System Model}
    \end{figure*}

In wireless communication, transmitted signals are processed by the wireless receiver to recover the original data. A traditional receiver consists of a series of interconnected modules, each performing a specific signal processing task. Figure \ref{fig:System Model} illustrates the architecture of a typical wireless communication system where the transmitter is consisted of source coding, encryption, channel coding, modulation, space-time coding and the receiver is consisted of key signal processing modules, including components for synchronization, channel estimation, equalization, space-time coding, demodulation, decryption, and decoding. The focus of this study is the receiver part of a wireless system.

The synchronization module of the receiver is responsible for the alignment of the receiver's timing and frequency with the incoming signal. Different research works have addressed the issue of improving the performance of synchronization, including the mobile station-assisted receiver-receiver approach in an urban environment for improved accuracy \cite{SyncRcvRCV}, frequency synchronization for precision \cite{SyncFreq}, and time synchronization algorithms such as timing-sync protocol for sensor networks (TPSN) and delay measurement time synchronization (DMTS) for wireless sensor networks (WSNs) \cite{SyncWSN}. Other works have also explored innovative methods like utilizing natural environmental signals for synchronization \cite{SyncNatural}, distributed synchronization \cite{Syncdistributed}, and GPS integration \cite{SyncGPS}. 

In wireless communication, the effect of the channel is characterized by shadowing, multipath fading, attenuation, interference, and noise. One of the important tasks in a wireless communication system is to estimate the channel state information (CSI). Althahab and Alrufaiaat highlight the importance of channel estimation for minimizing error rates in MIMO systems in \cite{chestimationsurvey}, while Yu and Gui in \cite{chestimationFIR} emphasize the necessity of channel estimation for reconstructing finite impulse response (FIR)  models in frequency-selective channels. Traditional methods like least squares (LS) and linear minimum mean square error (LMMSE) channel estimation \cite{ChannelEstimatonLMMSE} are complemented by modern approaches like DL \cite{CHestimationML5G} for enhanced accuracy in dynamic environments, and compressive sensing for efficient estimation in sparse channels \cite{chestimationCompressive}. Hybrid methods, such as combining Kalman filtering with compressive sensing, offer robustness in varying conditions \cite{ChestimationINM}. Ongoing research explores adaptive algorithms \cite{ChestimationOrthoMIMO} and DL \cite{ChestimationDL} to further improve CSI estimation accuracy and efficiency.

After channel estimation, restoring the channel integrity by compensating for the channel effects becomes necessary. Equalization deals with the adverse effects of channel distortion. Adaptive algorithms can outperform traditional algorithms like  least mean square algorithm (LMS) and constant modulus blind equalization algorithm (CMA) by effectively mitigating inter symbol interference (ISI) in fading channels \cite{equalizationAdaptiveKovac}, \cite{equalizationAdaptiveXi}, while more advanced methods like deep belief networks (DBNs) handle complex channel conditions \cite{equalizationDBN}. The choice of equalization technique depends on factors such as the specific wireless standard, application, and complexity considerations \cite{EqualizationMMSE-DFE}, \cite{EqualizationLinear}. In \cite{equalizationParityCheck}, the authors explore a new approach by integrating equalization with channel coding for enhanced robustness in mobile environments.

Space-time decoding is a crucial technique used in the receivers of a MIMO system. In a MIMO system, multiple antennas are used in both the transmitter and receiver, which allows the receiver to improve the reliability of the communication by increasing diversity gain. In addition, this approach can improve data transmission rates. The fundamental goal of space-time coding is to maximize diversity gain, encoding gain, and overall capacity while ensuring that the decoding process remains computationally efficient \cite{SpacetimecodingMimo}. There are different approaches to improving the performance of space-time coding and decoding. Cui et al. have proposed programmable coding metasurfaces \cite{SpaceTimecodingMetasurface} to enhance the capabilities of space-time coding systems by modulating radiation patterns and distributing information across various spatial angles. With the advancement of MIMO diversity and integration of space-time coding, the potential to substantially improve the overall system performance is high in adverse environments characterized by multipath propagation \cite{SpaceTimecodingEvolutio}. 

From the modulated carrier, the baseband signal is extracted by the receiver's demodulation module. The efficiency and accuracy of demodulation directly impact the overall performance of communication systems. The performance of the demodulator becomes more critical in challenging wireless environments characterized by noise and interference. There are broadly two types of demodulation techniques: coherent and non-coherent demodulation methods. These two methods have their own advantages and applications, which are discussed in \cite{DemodulationCNN}. Coherent demodulation relies on phase synchronization between the transmitter and receiver, which makes it a suitor for systems where accurate phase information can be maintained. In the quadrature amplitude modulation (QAM) modulation scheme, the demodulator uses the phase information to recover the transmitted symbols accurately \cite{DemodulationDL2019}. Recent advancements in DL have further improved coherent demodulation techniques by enabling the development of deep convolutional network demodulators that can effectively reduce bit error rates (BER) in complex environments \cite{DemodulationDL2019,DemodulationDL22}. On the other hand, non-coherent demodulation does not require phase synchronization.  In scenarios where maintaining phase information is challenging, non-coherent demodulation is advantageous \cite{DemodulationCNN}. 

The function of channel decoding module is recovering the transmitted information from the received signal. Traditional methods, such as maximum likelihood decoding and soft-decision decoding, have been widely used in conjunction with various coding schemes, including low-density parity-check (LDPC) codes and turbo codes. These methods aim to minimize the BER and enhance the overall reliability of communication systems \cite{ChannelcodingDL2019},\cite{Channelcoding2012}. In high-rate transmission scenarios, the complexity of these decoding algorithms can be a limiting factor. Recent research has explored the use of neural networks to perform channel decoding, leveraging their ability to learn complex mappings from received signals to transmitted symbols. For instance, DL-based approaches have shown promise in achieving performance comparable to traditional decoding methods while potentially reducing computational complexity \cite{ChannelcodingDL2019}, \cite{ChannelcodingDLMimo17}.

Source decoding, on the other hand, involves the reconstruction of the original information from the encoded data. In modern communication systems, this process is often integrated with channel decoding to form a unified decoding framework. Autoencoders, a type of DNN architecture, have been particularly effective in this context. They can be trained to perform both encoding and decoding tasks simultaneously. This makes it possible for the the communication system to perform end-to-end optimization \cite{Sourcecoding2020},\cite{SourcecodingAutoencoders2021}.

Each module in a traditional receiver faces unique challenges. Synchronization can be affected by noise, interference, and clock offsets. Channel estimation is often complex due to the dynamic nature of wireless channels and the presence of multipath propagation. Equalization algorithms may struggle to compensate for severe channel distortions, and demodulation can be impaired by noise and interference. Decoding errors can arise from channel impairments or limitations in the decoding algorithm.

It is important to note that errors in each module can propagate to subsequent modules, potentially amplifying the overall error rate. For example, an error in channel estimation can lead to incorrect equalization, which in turn can cause demodulation and decoding errors. DL-based receivers can help mitigate this error propagation by improving the accuracy and robustness of individual modules and by learning to compensate for errors in previous stages. The Authors in \cite{Zheng_2021} introduced a receiver model, \textbf{DeepReceiver}, that is able to replace traditional receiver modules such as carrier and symbol synchronization, channel estimation, equalization, demodulation and decoding. The proposed model eliminates the possibility of error propagation from one module to another by replacing the individual modules with a deep neural network module. It features a specialized network design, the one-dimensional convolution DenseNet (1D-Conv-DenseNet), which is adept at processing signals of varying lengths. A significant aspect of DeepReceiver is its ability to work with different types of signal encoding methods (modulation and coding schemes) without needing to know in advance which type is being used.

The data driven approach of DL makes it a high potential solution to address the challenges of traditional wireless receiver design. DL models can learn complex patterns and relationships in wireless signals. DL models can be trained to outperform traditional methods in various wireless communication scenarios by replacing receiver modules, such as synchronization, channel estimation, equalization, demodulation, and decoding.

\section{Deep Neural Networks Architectures}
\label{sec:Deep Neural Networks Architectures}

Deep neural networks represent a class of machine learning models characterized by multiple processing layers composed of interconnected neurons \cite{Deep_Learning}. The structure and function of these interconnected layers of neurons are similar to those of a human brain. The human brain-like structure uses a backpropagation algorithm, which enables it to learn intricate patterns from large representations of data. This section discusses different types of DNNs based on their architectures and functional properties. Some of the most commonly used types include MLPs, which are suited for structured data analysis; CNNs, which specialize in processing spatially structured data such as images and signals; RNNs, which are preferable to handle sequential and time-dependent data; GANs, which can generate synthetic data by learning from real samples; and autoencoders, which are commonly used for data compression and reconstruction. Furthermore, we will delve into the use cases of these models in the wireless communication domain in depth.

\subsection{Multilayer Perceptron (MLP)}

\begin{figure}[h]
    \centering
    \includegraphics[width=1\linewidth]{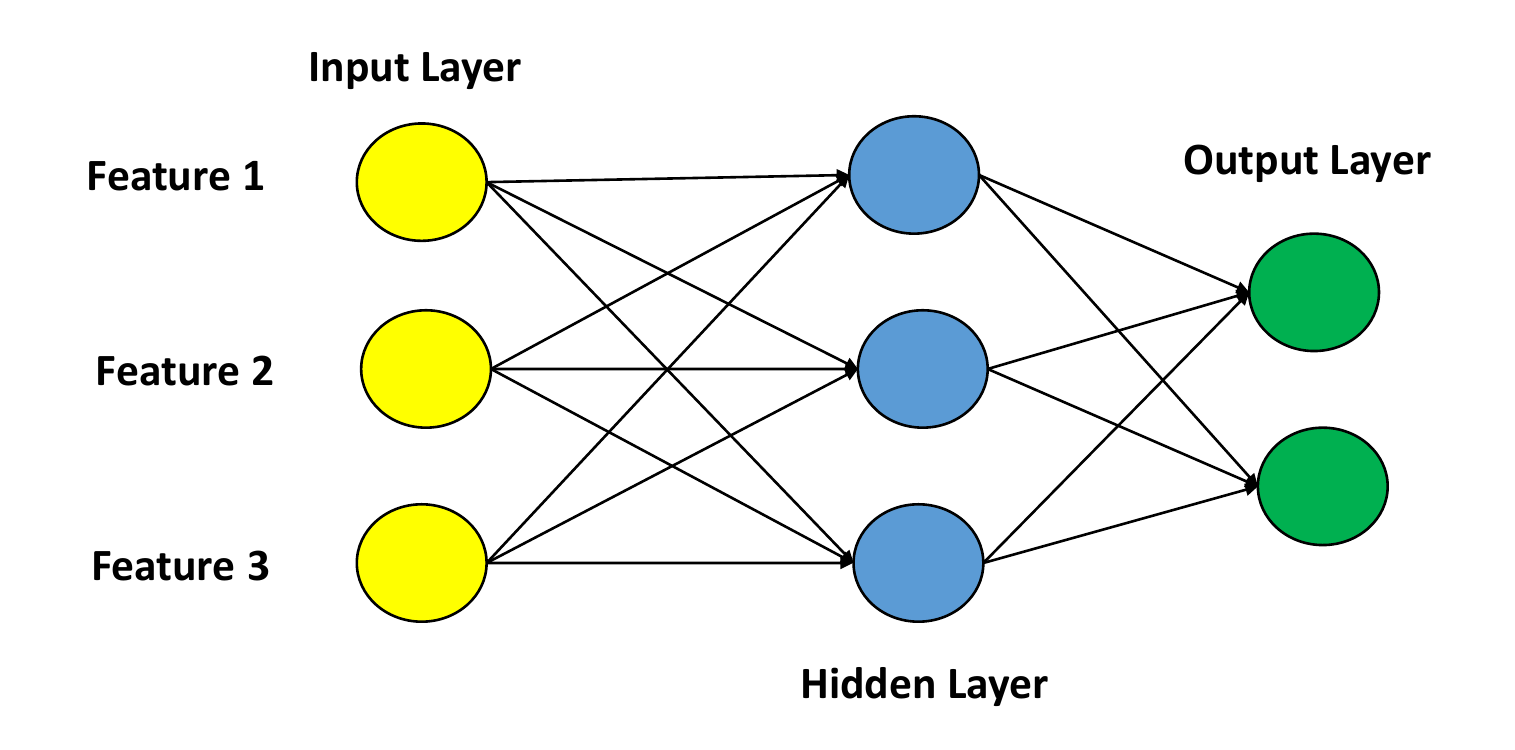}
    \caption{Structure of a MLP}
    \label{fig:Structure of a MLP}
\end{figure}

MLP is the most basic form of DNN. It consists of feedforward fully connected (each neuron of a layer is connected with every neuron of an adjacent layer) layers. Each neuron of MLP use nonlinear activation function which helps the network to find complex patterns from a given structured dataset. A simple structure of MLP with one input layer, one hidden layer, and one output layer is shown in Figure \ref{fig:Structure of a MLP}. Because of their simplicity, MLPs are useful in real life applications where low computation power is of concern. For detecting intrusion in WSNs, where the edge nodes have less computation power, authors in \cite{MLPwsn} have shown that MLP has demonstrated higher accuracy in classification compared with state of the art models while dealing with real time diverse non-linear massive data. The issue of localization in internet of everything (IoE) has been addressed in \cite{MLPmmw} using MLP based model.

\subsubsection*{MLPs in Deep Learning-Based Receivers}

MLPs are employed in DL-based receivers for their ability to model and predict complex relationships in data:

\begin{itemize}
    \item \textbf{Channel Estimation:} They can estimate the channel effects to improve the accuracy of received signal interpretation \cite{CE_MLP}.
    
    \item \textbf{Distortion correction:} MLPs can improve the quality of the received signal by effectively mitigating signal nonlinear distortions \cite{Distortion_MLP} .
    
    \item \textbf{Signal Classification:} Useful in classifying modulation schemes or identifying signal types based on their characteristics \cite{AMC_MLP}.
    
    \item \textbf{Resource Allocation:} They can optimize resource allocation by predicting network traffic and user demand \cite{ResourceAllocation_MLP}.
\end{itemize}

\subsection{Convolutional Neural Network (CNN)}

\begin{figure}[h]
    \centering
    \includegraphics[width=1\linewidth]{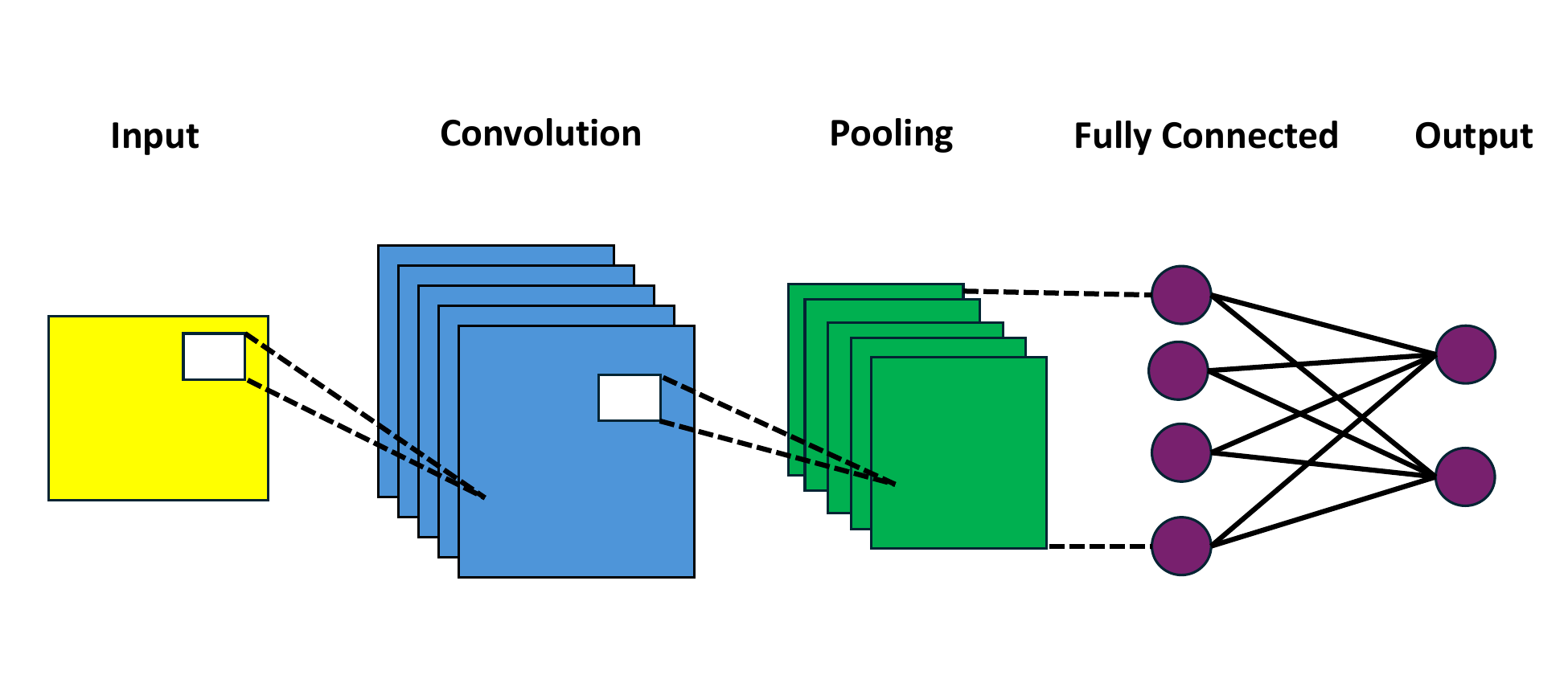}
    \caption{Architecture of a Convolutional Neural Network}
    \label{fig:Structure of a CNN}
\end{figure}

 The architecture of a basic CNN (shown in Figure \ref{fig:Structure of a CNN}) consists of the input layer, convolutional layers, pooling layers, a fully connected layer, and an output layer. The convolutional layers, which are the key layers of a CNN architecture, use filters or kernels for feature maps. The pooling layers are used to downsample the feature maps, i.e., these layers reduce the spatial dimension. After convolutional and pooling layers, high-level reasoning is performed in fully connected layers. Although CNNs are mainly popular for their application in image and video processing because of their ability to learn intricate patterns from grid-like data, CNNs are widely used in the wireless communication domain. An intelligent receiver, \textbf{CNN-IR}, is proposed in \cite{article1} that utilizes a CNN to minimize data recovery errors caused by channel impairments like multipath fading and noise. It replaces traditional channel estimation and equalization techniques with a trained CNN that learns the complex relationship between transmitted and received signals. CNNs have been used for channel estimation in \cite{CNNCE}, \cite{CNNCEBeamspace}, \cite{cnnchannelestimationofdm}, for spectrum sensing in \cite{CNNSpecSensing}, and for interference identification in \cite{CNNinterference}.
 
\subsubsection*{CNNs in Deep Learning-Based Receivers}

In wireless communication systems, CNNs can be used innovatively in DL-based receivers:

\begin{itemize}
    \item \textbf{Channel Estimation:} CNNs can improve channel estimation accuracy across various communication scenarios, such as massive MIMO \cite{CE_CNN_MASSIVEMIMO} and reconfigurable intelligent surface (RIS)-assisted communications \cite{CE_CNN_RIS}.
    
    \item \textbf{Signal Demodulation:} CNNs can be utilized to process received signal data, enabling robust demodulation in complex environments characterized by multipath propagation and interference \cite{Demod_CNN}.
    
    \item \textbf{Signal Decoding:} CNNs can be effectively employed in decoding processes of H.265 encoded videos \cite{Decod_Vid_CNN}. It can also be used in conjunction with belief propagation decoding for polar codes \cite{Decod_PolarCodes_CNN}. 

    
    \item \textbf{Feature Extraction:} CNNs excel in extracting relevant features from signals which is crucial in semantic communication.
\end{itemize}

\subsection{MobileNet}

MobileNet is a specialized class of CNN architecture that gained popularity because of its ability to be implemented in devices with limited processing powers. MobileNet is an efficient model designed for mobile and embedded vision applications \cite{MobileNet}. The main innovative feature of this architecture is the use of depth-wise separable convolutions. In this architecture, the convolution process is divided into two convolutional processes. The first one is a depth-wise convolution, followed by the second part, a point-wise convolution. This separation in the convolutional process helps to reduce the computational complexity and the size of the model.

\subsubsection*{MobileNet in Deep Learning-Based Receivers}

MobileNet is primarily well-suited for mobile and edge computing environments. However, its efficiency and adaptability make it suitable choice for deployment in DL-based receivers:

\begin{itemize}
    \item \textbf{Automatic Modulation Classification:}  The ability of MobileNet to extract relevant features from received signals can improve classification accuracy \cite{MobiNet_2023}.
    
    \item \textbf{On-device Learning:}  MobileNet can also be utilized for channel estimation in wireless communication systems \cite{MobiNet_2023}.
    
    \item \textbf{Energy Efficiency:} Its low computational requirement makes it a preferred choice for battery-powered devices.
    
    \item \textbf{Improved Response Time:}  MobileNet's lightweight architecture is ideal for applications requiring quick response times, such as intelligent pick-and-place systems, which can be analogous to tasks in wireless communication \cite{Response_MOBI}.
\end{itemize}

\begin{table*}[htbp]
\centering
\caption{Distinctive Use Cases and Benefits of Deep Neural Network Architectures}
\label{table:nn_use_cases}
\begin{tabular}{|m{4cm}|m{8.5cm}|m{4cm}|}
\hline
\textbf{DNN Architecture} & \textbf{Distinctive Benefits} & \textbf{Use Cases} \\
\hline
MLPs (Multilayer Perceptrons) & Suitable for robust mapping between inputs and outputs, enhancing receiver performance in non-ideal channel conditions through improved channel estimation and signal detection. & Channel Estimation, Signal Detection, Decoding \\
\hline
CNNs (Convolutional Neural Networks) & Excels in spatial data processing, facilitating accurate channel estimation, signal detection, and modulation classification by capturing spatial hierarchies in data, improving performance in environments with interference and signal distortion. & Channel Estimation, Signal Detection, Modulation Classification \\
\hline
MobileNet & Optimized for efficient computation on mobile devices, improving energy efficiency and computational speed in receivers for tasks like signal classification and interference analysis, crucial for adaptive and real-time communication systems. & Signal Classification, Interference Analysis \\
\hline
RNNs (Recurrent Neural Networks) & Ideal for sequential data processing, improving error correction and dynamic adaptation in receivers by effectively handling time series data in changing channel conditions. & Error Correction, Dynamic Adaptation \\
\hline
GANs (Generative Adversarial Networks) & Powerful for generating synthetic data, augmenting datasets for training DL models in wireless communication, and simulating channel effects for improved receiver design.  & Data Augmentation, Channel Simulation \\

\hline
Autoencoders & Jointly optimize transmitter and receiver, replacing traditional modulation, equalization, and demodulation methods; handle different mapping schemes and channel conditions (AWGN, fading, non-Gaussian noise).  & End-to-End Communication System Optimization, Channel Agnostic Learning \\ 

\hline
\end{tabular}
\end{table*}

\subsection{Recurrent Neural Network (RNN)}

\begin{figure}[h]
        \centering
        \includegraphics[width=.99\linewidth]{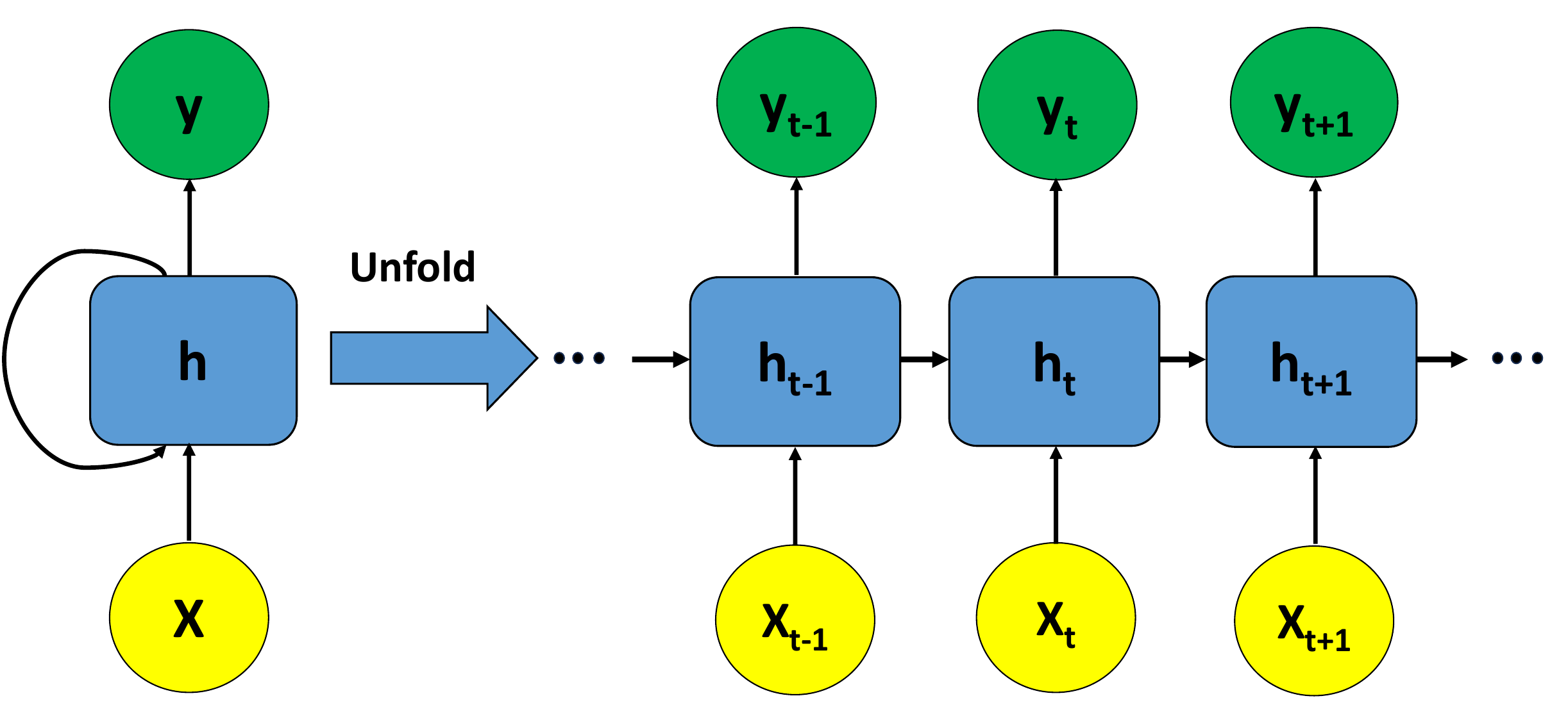}
        \caption{Basic Recurrent Neural Network}
        \label{fig:RNN Unfolded}
    \end{figure}

It is essential to retain the information of the previous input
state for sequential data-related tasks like time series analysis,
natural language processing, or any other tasks that involve
temporal data. However, feed-forward neural networks like
MLP and CNN, where data from one layer to another layer moves unidirectionally, do not have this ability to remember the input information of the previous state. In a RNN, the output of the previous step is used as an input. The recurrent unit, which has a hidden state that essentially works as memory, is the fundamental block of RNN Figure \ref{fig:RNN Unfolded}. One of the key differences between feed-forward neural network and RNN is that feed-forward neural network does not have any looping node. We can express an RNN as a function $f_{\theta}$ of type:

\begin{equation}
(x_t, h_t) \xrightarrow{f_{\theta}} (y_t, h_{t+1})
\label{eq:rnn_function}
\end{equation}
where

\begin{itemize}
\item $x_t$: input vector;
\item $h_t$: hidden vector;
\item $y_t$: output vector;
\item $\theta$: neural network parameters.
\end{itemize}

Many research works are done in wireless communication by integrating RNN to analyze sequential data. For estimating CSI, the authors in \cite{RNN5GCE} have demonstrated that their proposed online pilot-assisted estimator, which uses RNN, shows robustness and better performance than conventional estimators. The authors of \cite{RNNCoChanIn} have used RNN architecture to address the issue of co-channel interference. In \cite{RNNOFDMIM}, an RNN-based signal detection scheme for OFDM with index modulation is proposed. This signal detector can effectively extract features by storing information of prior time steps for a long time. The authors in \cite{RNNMassive} have demonstrated that their proposed RNN-based channel predictor for Massive MIMO system outperforms conventional system with very low complexities. 

\subsubsection*{RNNs in Deep Learning-Based Receivers}
RNNs find application in DL-based receivers within wireless communication systems for their unique ability to process sequential data:

\begin{itemize}
    \item \textbf{Channel Estimation:} RNNs can be useful for channel estimation by analyzing sequences of received signals \cite{CE_RNN}. 

    \item \textbf{Channel Decoding:} RNNs are effectively employed in channel coding and decoding processes \cite{CH_DEC_RNN_2022}, \cite{CH_DEC_RNN_kim2018}. RNNs can adaptively learn the underlying structure of the data, enabling more efficient decoding in the presence of noise \cite{CH_DEC_RNN_2018}.

    \item \textbf{Modulation Recognition:} RNNs can recognize modulation patterns in received signals with varying inpur sequence length \cite{AMC_RNN}.

    \item \textbf{Inter symbol Interference Analysis:} By processing real-time signal, RNNs can mitigate inter symbol interference (ISI) in broadcasting systems \cite{ISI_RNN}.

    \item \textbf{Demodulation:} RNNs can be employed to enhance demodulation processes across various modulation schemes, including Binary Phase Shift Keying (BPSK) \cite{DemodulationDL22}  and Amplitude Shift Keying (ASK) \cite{Demod_RNN_ASK}.
\end{itemize}

\subsection{Generative Adversarial Networks (GANs)}

\begin{figure}[h]
    \centering
    \includegraphics[width=1 \linewidth]{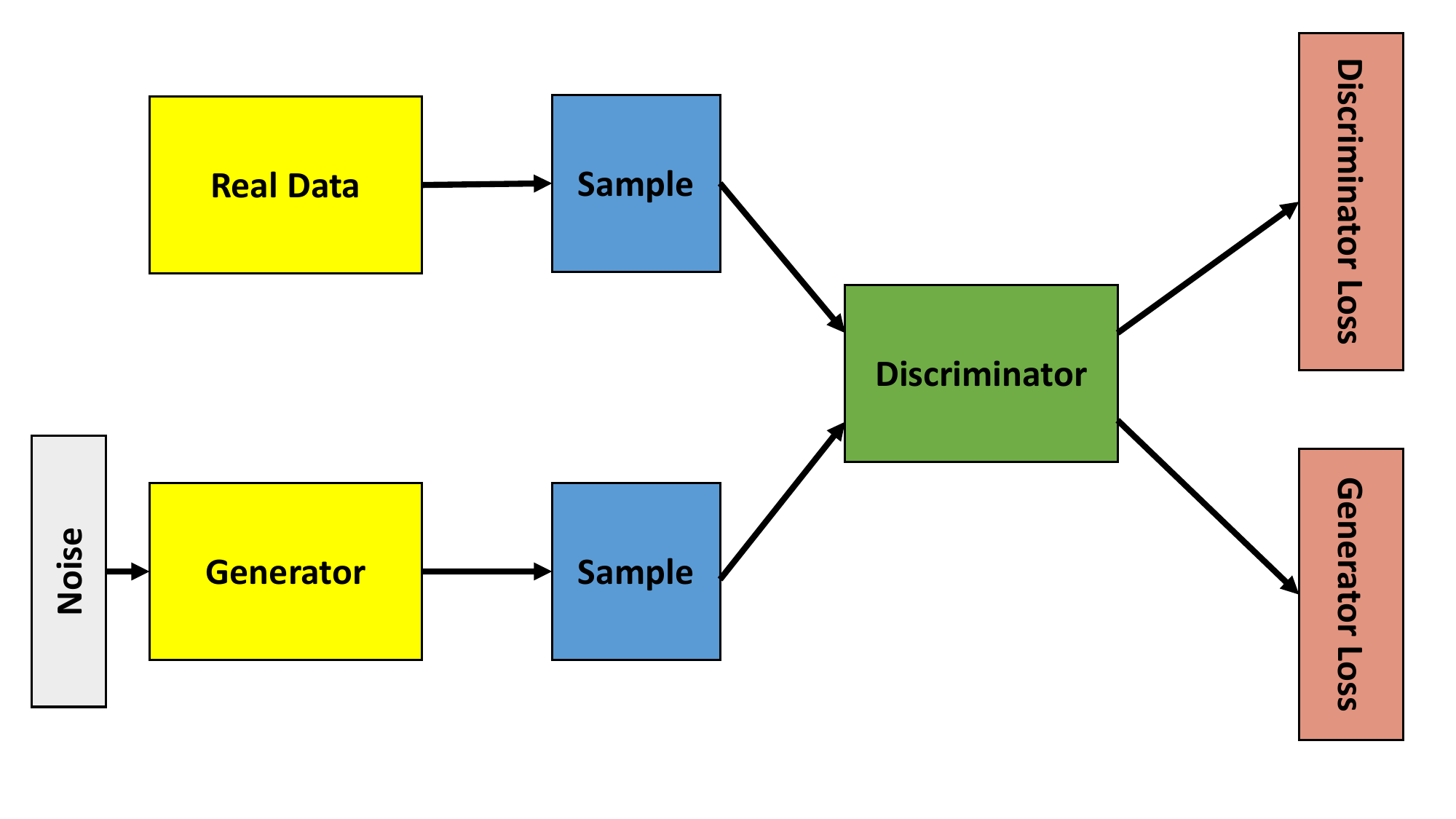}
    \caption{Structure of a GAN}
    \label{fig:GAN}
\end{figure}

In 2014, Ian Goodfellow and his colleagues proposed a new type of DL Model known as generative adversarial networks (GANs) \cite{generativeadversarialnetworks}. There are two neural network components in a GAN: a generator and a discriminator. The structure of a GAN is shown in Figure \ref{fig:GAN}. The two networks of GAN engage in a competitive zero-sum game. The generator is trained to produce data samples that mimic genuine data; on the other hand, the discriminator tries to differentiate between actual data and created samples. The two networks of GANs are:

\textbf{Generator (G):} This network takes random noise as input and generates data samples as output. The purpose of the generator is to create samples similar to real data so that the created data samples can not be distinguished from the real data. 

\textbf{Discriminator (D):} This network takes both authentic data samples and created samples from the generator as input, aiming to categorize them accurately. The objective of the discriminator is to distinguish between authentic and counterfeit samples.

 The model of a GAN can be mathematically expressed as a min-max game between the generator and the discriminator. The generator aims to minimize the loss function, and the discriminator aims to maximize the same loss function. The Loss function can be expressed as : \\

\begin{align}
\min_G \max_D V(D,G) &= \mathbb{E}_{x \sim p_{data}(x)}[\log D(x)] \nonumber \\ 
&\quad + \mathbb{E}_{z \sim p_z(z)}[\log(1 - D(G(z)))] 
\label{eq:gan_loss}
\end{align}
where:

\begin{itemize}
\item $x$ represents a real data sample.
\item $z$ represents a random noise sample.
\item $p_{data}(x)$ is the distribution of the real data.
\item $p_z(z)$ is the distribution of the random noise.
\item $G(z)$ is the output of the generator.
\item $D(x)$ is the output of the discriminator.
\end{itemize}

Gan can learn from limited data, which is particularly advantageous in wireless communication, where there is a challenge of training with high-quality real-time labeled data. GAN is a useful tool for data augmentation for training cognitive radio with the task of spectrum sensing \cite{GANsprectrumsensing}, automatic modulation classification (AMC) \cite{GANAMC} and detection of anomalous signals. \cite{GANRadioanamoly}. Hao et al. integrated a GAN model in their end-to-end DL-based communication system to represent the effects of channels \cite{GANDLEndtoEnd}.

\subsection{Autoencoders}

Autoencoders are unsupervised learning models that can compress and rebuild input data \cite{Autoencodesurvey}. They possess an encoder that translates input data into a low-dimensional representation and a decoder that reconstructs the original input data from the low-dimensional representation. Autoencoders can be mathematically represented as follows :

\begin{itemize}
\item \textbf{Encoder:}  $h = f(x)$ 
\item \textbf{Decoder:} $x' = g(h)$
\end{itemize}

where:

\begin{itemize}
\item $x$ is the input data.
\item $h$ is the encoded representation (latent code) of the input data.
\item $x'$ is the reconstructed version of the input data.
\item $f$ is the encoder function.
\item $g$ is the decoder function.
\end{itemize}

 The encoder part converts the input data into lower-dimensional latent code. The encoder usually has several layers of neurons, including fully connected and convolutional layers. Each layer performs a non-linear transformation over the previous layer's output. The decoder component of the encoder takes the latent code as input and gives an approximation of the original data as output. It is similar in design to the encoder, with layers that apply the inverse operations of the corresponding encoder layers.

In wireless communication, autoencoders have gained significant interest among researchers because of their ability to facilitate end-to-end learning and optimization of communication systems \cite{AutoencodrPhysicalLayer},\cite{AuoencoderTransciever}. Researchers have created the advanced method of joint source-channel coding (JSCC) by adding autoencoders to wireless communication systems. This method eliminates the need for separate coding and modulation stages. This approach has demonstrated a more efficient mapping of source signals directly to channel inputs \cite{AutoendoerJsccImage}, \cite{AutoencoderJsccBER}. Furthermore, autoencoders have been used to improve the robustness and efficiency of wireless communication systems by learning optimal coding and decoding strategies directly from data \cite{DLforWireless}.\\

Table \ref{table:nn_use_cases} summarizes the distinctive benefits and use cases of the DNN architectures analyzed in this section.

\section{Application of Deep Learning based receivers}
\label{Sec: Application of Deep Learning based receivers}

Many research works have exploited the implementation of DL in wireless receivers across different applications. In this section, we have looked into the different domains and technologies where DL-based receiver is used, including joint optimization of transmitter and receiver, semantic communication, task-oriented communication, OFDM, MIMO, Next-G Wireless, successive interference cancellation (SIC), and soft interference cancellation (SoftIC). 
\subsection{Deep Learning in Joint Optimization of Transmitter and Receiver}
One of the popular applications of DL is the joint optimization of the transmitter and receiver. This can be done by considering the communication system as an autoencoder built with different neural networks \cite{AuoencoderTransciever,TransciverAutoencoder2017,TranscieverAutoencoder2016,AutoencodetTransciverCNN}. The authors in \cite{AutoencodetTransciverCNN} introduce a convolutional autoencoder neural network structure that jointly optimizes the transmitter and receiver, replacing traditional modulation, equalization, and demodulation methods. The proposed structure can handle different levels of mapping schemes under various channels, including AWGN, fading channels, and non-Gaussian noise channels. Additionally, The authors have also presented a comparison between different machine learning techniques used for building an autoencoder which is shown in Table \ref{Machine Learning Methods used in autoenoder}. Their study shows that convolutional autoencoders (CNN-AE) can incorporate channel design with low complexity and variable input lengths. DNN-based autoencoders (DNN-AE) are more complex and unable to incorporate variable length input. On the other hand, autoencoders can be implemented with a boosted tree which has low complexities but is unable to incorporate variable length input and channel design. \\

\begin{table}[!h]
\centering
\caption{ Properties of Different Machine Learning Methods used in autoenoder}
\label{Machine Learning Methods used in autoenoder}
\begin{tabular}{|p{1.5CM}|p{1.5CM}|p{1.7CM}|p{1.8CM}|}
\hline
\textbf{Methods} & \textbf{Incorporate Channel} & \textbf{Low Complexity} & \textbf{Variable Input Length} \\
\hline
Boosted Tree & $\times$ & $\checkmark$ & $\times$ \\
\hline
DNN-AE & $\checkmark$ & $\times$ & $\times$ \\
\hline
CNN-AE & $\checkmark$ & $\checkmark$ & $\checkmark$ \\
\hline
\end{tabular}
\end{table}

It is a challenging task to estimate the channel for jointly training the transmitter and the receiver. The challenge is that the backpropagation algorithm used in training DNNs requires a differentiable channel function. This means that the function describing the channel's behaviour needs to be expressed in a way that makes it possible to compute gradients. However, real-world channels often exhibit complex, non-linear characteristics that are difficult to model in a differentiable form. Researchers have explored a range of techniques to address this challenge. The most popular approach is to use GANs and their variants conditional GAN (cGAN) \cite{ChannelAgnosticEnd-to-End}, variational GAN \cite{approximatingvoidlearningstochastic}, wasserstein GAN (WGAN) \cite{WGAN-BASEDtraining}. The authors in \cite{s24102993} propose a mixture density network (MDN) to train to approximate the channel distribution at the receiver. Unlike GAN, two different models do not need to be trained in this approach, which simplifies the training process and makes it more suitable for energy-constrained devices like those in the IoT. A comparison of different methods used for training the encoders used in joint optimization of transmitter and receiver is shown in Table \ref{tab:channel_estimation_comparison}.

\begin{table}[!h]
\centering
\caption{Comparison of Different Channel Estimation Methods for Encoder Training}
\label{tab:channel_estimation_comparison}
\begin{tabular}{|p{.6cm}|p{2.1cm}|p{4.2cm}|}
\hline
\textbf{Paper} & \textbf{Channel Estimation Method} & \textbf{Unique Contribution} \\ \hline
\cite{ChannelAgnosticEnd-to-End} & Conditional GAN  & Leverages GANs for channel estimation, enabling channel-agnostic learning and the joint optimization of transmitters and receivers in an end-to-end approach.\\ \hline
\cite{approximatingvoidlearningstochastic}  & Variational GAN & Focuses on learning the probability distribution of the channel to capture complex, stochastic behaviors (including non-linear and non-Gaussian effects). \\ \hline
\cite{WGAN-BASEDtraining}& Wasserstein GAN &  Improves training stability and learned an implicit channel model, which simplifies autoencoder training by removing the need for continuous feedback. \\ \hline

\cite{s24102993}& Mixture Density Network & Reduces the complexity of training process with maintaining similar performance.  \\ \hline
\end{tabular}
\end{table}

\subsection{Deep Learning in Semantic Communication}

\begin{figure}[h]
    \centering
    \includegraphics[width=1 \linewidth]{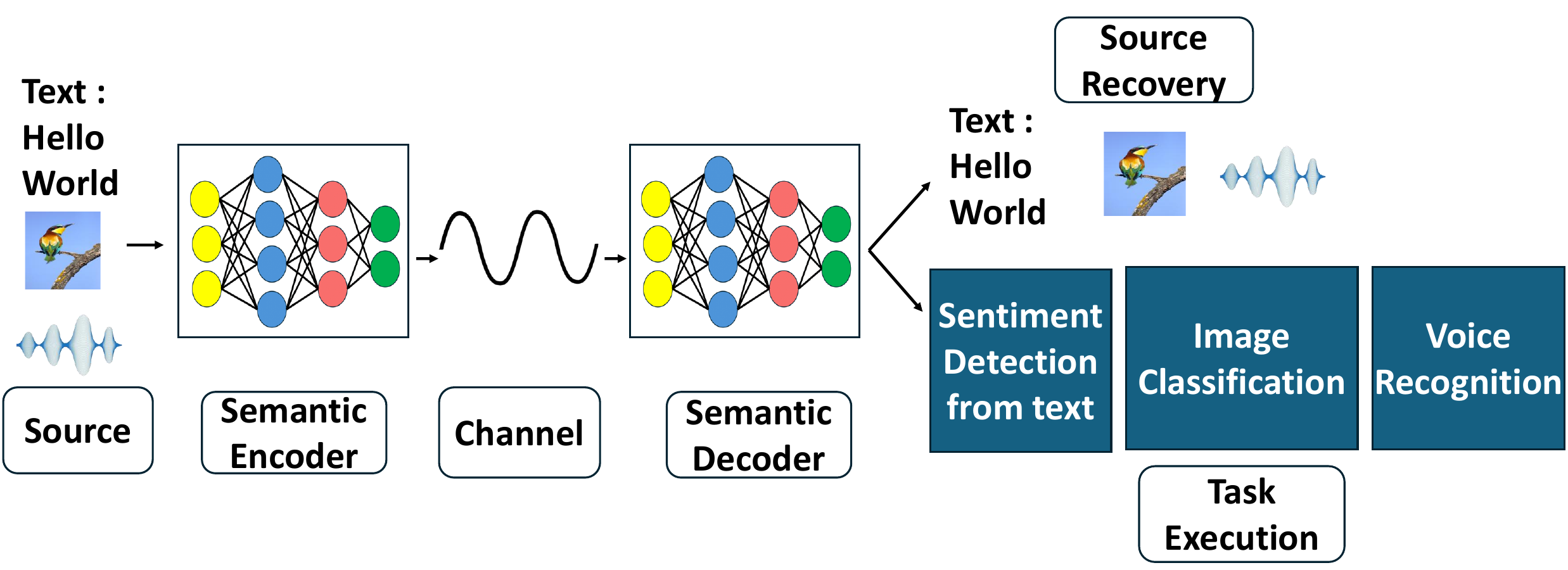}
    \caption{Schematic of a DL enabled Semantic Communication System}
    \label{fig:Schematic of a DL enabled Semantic Communication System}
\end{figure}

The principle of semantic communication is to quantify and understand the meaning of information in communication systems. The authors in \cite{qin2022semantic} contributed to defining the basic concepts of semantic theory, such as Semantic Information, Semantic Entropy, Semantic Channel, and Semantic Communication. Shannon's information theory focuses on the transmission of symbols or bits regardless of their meaning. Semantic communication focuses on the underlying meaning or semantics of the material rather than the raw data itself to express information. Instead of sending all data bits, it identifies and prioritizes the most relevant and meaningful features of the content. In recent years, numerous intriguing works for semantic communications have been developed, largely due to developments in DL. Authors in \cite{beyond} presented a schematic diagram of a semantic communication system leveraging DL-based encoder and decoder, which is shown in Figure \ref{fig:Schematic of a DL enabled Semantic Communication System}. Semantic communication performs better under low signal-to-noise ratio (SNR), high interference, and high fading conditions for reconstructing the original data by leveraging the extracted semantic features. One of the key applications of extracting semantic features is that it can be used to complete a specific task at the receiver end, such as image classification, voice recognition, and sentiment detection. The task-oriented communication part is discussed in the following subsection. \\

The authors in \cite{Xie_2021} proposed the \textbf{DeepSC} system that uses the transformer architecture to extract and transmit the semantic meaning of text messages. In the DeepSC system, semantic and channel coding layers are jointly designed, and the system demonstrates superior performance compared to traditional communication systems in poor channel conditions. Improvement is shown in both the Bilingual Evaluation Understudy (BLEU) score and sentence similarity metrics by DeepSC. The BLEU score shows an 800 \% improvement over conventional methods at a 9 dB SNR. The result analysis of the sentence similarity metric highlights that the DeepSC can maintain high semantic fidelity even when the BLEU score is low (around 0.2 at 12 dB SNR).\\

\textbf{DeepSC-ST}, a DL-enabled semantic communication system, is introduced for efficient speech transmission in \cite{DeepSC-ST}. In contrast to conventional systems that struggle to deal with limited spectrum resources and dynamic channel conditions, DeepSC-ST substantially decreases the data required for transmission and performs better under fluctuating channel conditions, particularly in low SNR settings. The model extracts semantic properties related to text from the speech input, transfers these compressed features, and subsequently reconstructs the speech at the receiver utilizing both the received features and the speaker's identification. The proposed model substantially decreases the data required for transmission and performs better under fluctuating channel conditions, particularly in low SNR settings. The receiver of DeepSC-ST consists of two elements: the channel decoder, constructed with three dense layers, processes the received symbols and translates them into text-related semantic properties, and the feature decoder, which transforms the semantic information into a comprehensible text transcription.

 \textbf{DeepWive}, reinforcement learning enabled end-to-end framework for video transmission with variable bandwidth, has been developed in \cite{DeepWive}. The paper demonstrates that DeepWiVe outperforms traditional separation-based approaches, such as H.264/H.265 video compression combined with LDPC channel codes, in various channel conditions. Furthermore,  a JSCC scheme for video transmission over the air with to minimize the end-to-end transmission rate-distortion has been designed by the authors in \cite{wang2022wirelessdeepvideosemantic}. The paper proposes the Deep Video Semantic Transmission (\textbf{DVST}) framework, which integrates nonlinear transform and deep joint source-channel coding to enable end-to-end video transmission over wireless channels. DVST requires only 40 \% to 80\% of the channel bandwidth compared to H.264 + LDPC to achieve the same reconstruction quality in terms of PSNR, implying significant bandwidth savings. 

With a novel semantic error detector, the authors of \cite{semanticvideoconference} presented a semantic transmission strategy for video conferencing. The speaker's photo is provided as a prior information which helps to reconstruct the speaker's facial expression movement. The devised plan significantly reduces the need for wireless resources. A comparison of DL-based semantic
communication techniques and the contributions of individual works is presented in Table \ref{tab:semantic_communication_comparison}.

\begin{table}[!h]
\caption{Comparison of Deep Learning-based Semantic Communication Techniques}
\label{tab:semantic_communication_comparison}
\begin{tabular}{|p{.7cm}|p{2.85 cm}|p{4cm}|}
\hline
\textbf{Paper} & \textbf{Technique} & \textbf{Unique Contribution} \\ \hline
 \cite{Xie_2021} & Transformer-based joint semantic and channel coding & Achieves significant improvement in semantic fidelity for text transmission, particularly in poor channel conditions, as evidenced by BLEU score and sentence similarity metrics.  \\ \hline
 \cite{DeepSC-ST} &  Speech-to-text semantic feature extraction and transmission with speaker identity-aided reconstruction & Reduces data transmission requirements and improves performance in fluctuating channel conditions by transmitting compressed semantic features instead of raw speech data. \\ \hline
 \cite{DeepWive} & Reinforcement learning-enabled end-to-end framework for video transmission with variable bandwidth & Outperforms traditional separation-based approaches (H.264/H.265 + LDPC) in various channel conditions by adapting to variable bandwidth. \\ \hline
\cite{wang2022wirelessdeepvideosemantic} & Deep joint source-channel coding with nonlinear transform for video transmission &  Minimizes end-to-end transmission rate-distortion and achieves significant bandwidth savings compared to H.264 + LDPC while maintaining reconstruction quality.\\ \hline
\cite{semanticvideoconference} & Semantic transmission strategy with a novel semantic error detector and prior information (speaker's photo) & Reduces wireless resource requirements for video conferencing by prioritizing semantically relevant information and utilizing prior information. \\ \hline

\end{tabular}
\end{table}


\subsection{Deep Learning in Task-Oriented Communication}

Task-oriented communication is a style of communication focused primarily on achieving a specific goal or completing a task. It emphasizes efficiency, clarity, and directness in conveying information relevant to the task at hand. The focus is on the outcome, and the communication is often structured and organized.

\begin{figure*}[htbp]
    \centering
    \includegraphics[width=.99  \linewidth]{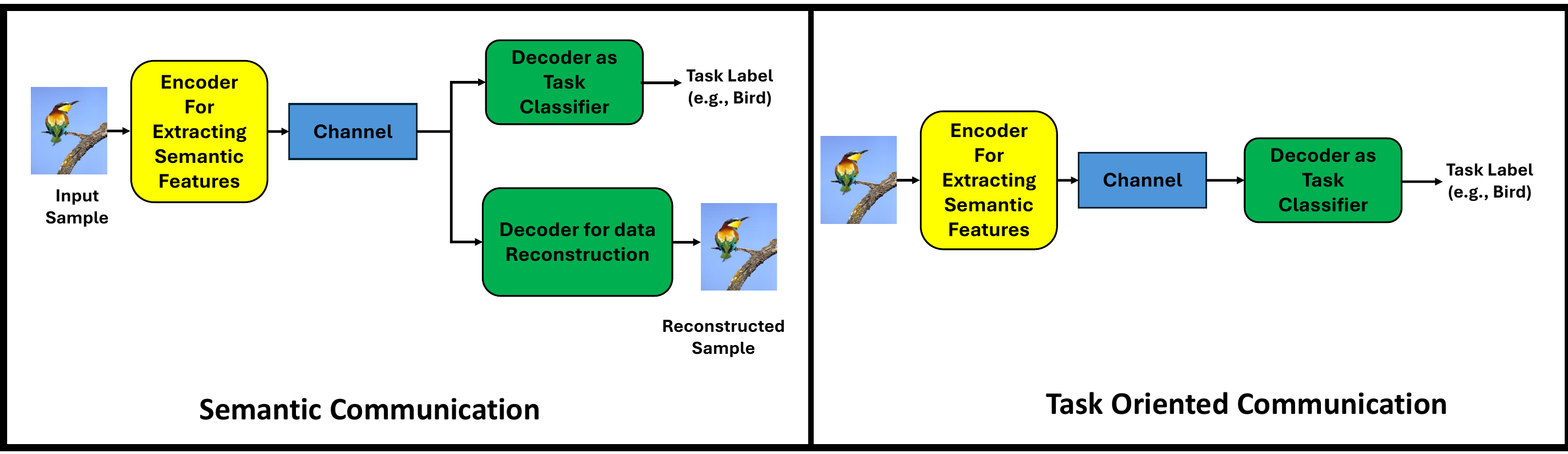}
    \caption{Comparison of Semantic Communication and Task-Oriented Communication}
    \label{fig:Comparison of Semantic Communication and Task Oriented Communication}
\end{figure*}

It is easy to get confused with semantic and task-oriented communication since semantic task-oriented communication uses semantic features to complete a specific task. Task-oriented communication can be considered as an application of Semantic Communication \cite{SecureSemantic}. A task-oriented grasping model is presented by the authors in \cite{tocrobotics}. Task-oriented communication in this context involves providing instructions to a robot to grasp an object based on its function or purpose (semantic affordance). For example, ``Grasp the handle of the mug" focuses on the specific task, while ``Grasp the mug so you can drink from it" incorporates semantic understanding. In semantic communication not only a specific task is executed at the receiver end but also the original data is reconstructed using the semantic features. In Figure \ref{fig:Comparison of Semantic Communication and Task Oriented Communication}, a comparison of semantic communication and task-oriented communication is presented. 

The information bottleneck (IB) method is a way to extract the most relevant information from a variable \(S\) (the observation) about another variable \(V\) (the relevant information) that we cannot directly observe \cite{SecureSemantic}. This is done by finding a mapping from \(S\) to a new representation \(U\) that maximizes the information about \(V\) while minimizing the information about \(S\) \cite{beyond}.

\subsubsection*{Mathematical Formulation}

The IB method is formulated as an optimization problem:

\begin{align}
L^{IB}_{\beta}(P_{S,V}) = \max_{P_{U|S}} I(U;V) - \beta I(U;S)
\end{align}

where:
\begin{itemize}
    \item \(P_{S,V}\) is the joint probability distribution of \(S\) and \(V\).
    \item \(P_{U|S}\) is the conditional probability distribution of \(U\) given \(S\), which represents the optimal mapping we want to find.
    \item \(I(U;V)\) is the mutual information between \(U\) and \(V\), measuring how much information \(U\) contains about \(V\) (relevance).
    \item \(I(U;S)\) is the mutual information between \(U\) and \(S\), measuring how much information \(U\) contains about \(S\) (complexity).
    \item \(\beta\) is a parameter that controls the trade-off between relevance and complexity.
\end{itemize}

The goal is to find the optimal mapping \(P_{U|S}\) that maximizes the relevance \(I(U;V)\) while minimizing the complexity \(I(U;S)\), which forms a markov chain \(U - S - V\).\\

The IB method is particularly useful for task-oriented compression, where the aim is to compress data for a specific task. The variable \(V\) represents the relevant information for the task, and the IB method finds the most compact representation \(U\) that retains enough information about \(V\) to perform the task accurately. This is different from traditional compression, which aims to minimize the distortion between the original and reconstructed data regardless of the task.  

The authors in \cite{beyond} emphasize that while information theory provides a theoretical foundation, practical applications often lack statistical information and require inferences based on single data samples. Machine learning, especially DL, offers a data-driven framework to address this challenge. Sagduyu et al. \cite{DeeplearingdrivenTOC} explore the use of DL to optimize task-oriented communications, focusing on the timeliness and accuracy of machine learning tasks executed at the receiver end. The concept of peak age of task information (PAoTI) is introduced to measure the freshness of information in task-oriented communication. The paper presents experiments with image classification tasks using datasets like MNIST and CIFAR-10 to demonstrate how encoder-decoder pairs can be optimized for both accuracy and latency. A dynamic update mechanism is proposed to adapt to varying channel conditions and arrival rates, further enhancing task performance.

\subsection{Deep Learning in OFDM Receiver}

Orthogonal frequency division multiplexing (OFDM) is a popular modulation technique in modern wireless communication. DL models like MLPs, RNNs, and CNNs are employed to replace conventional channel estimation, equalization, and demodulation modules, offering improved reliability and performance in adverse channel conditions.

\begin{figure}[h]
    \centering
    \includegraphics[width=1 \linewidth]{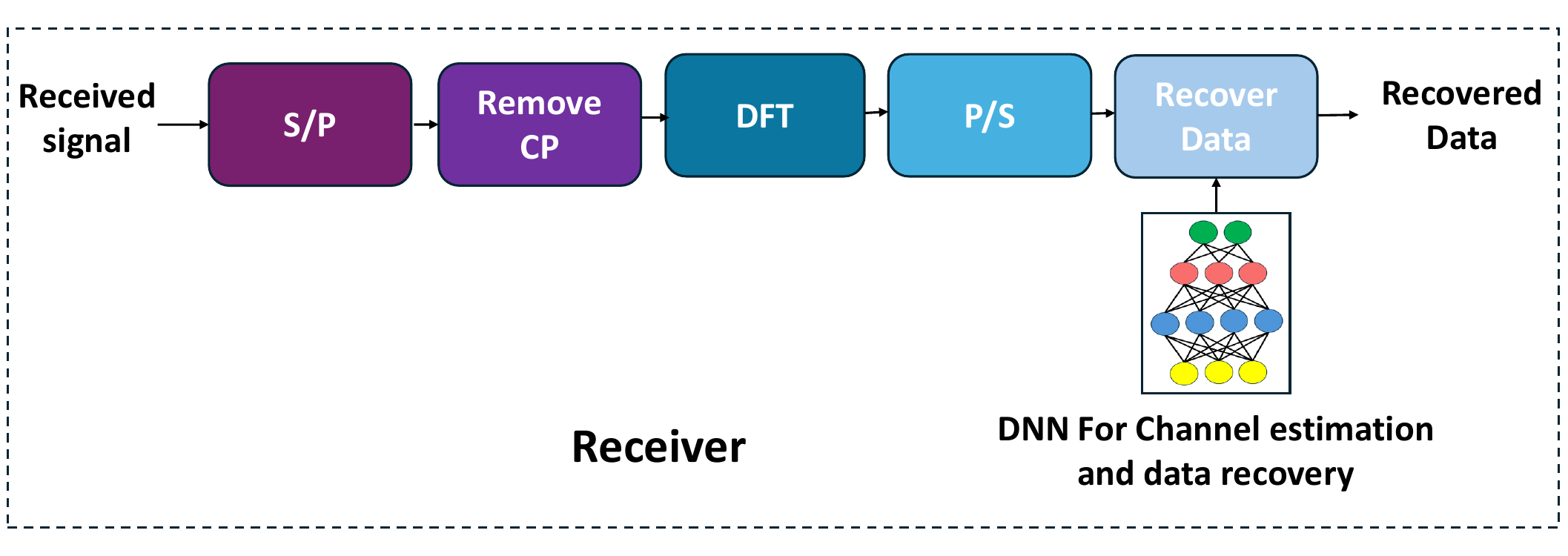}
    \caption{DNN bases OFDM Receiver of Channel Estimation and Data Recovery}
    \label{fig:DNN bases OFDM Receiver of Channel Estimation and Data Recovery}
\end{figure}

Several studies have explored the integration of DL into OFDM systems to enhance performance and address traditional limitations. The work in \cite{8052521} deploys DNNs for implicit CSI estimation and direct symbol recovery, which is shown in Figure \ref{fig:DNN bases OFDM Receiver of Channel Estimation and Data Recovery}. The authors trained their model in diverse channel conditions, and their model demonstrates better reliability for missing pilots, CP removal, and nonlinear distortions. The result showed better performance compared with traditional minimum mean square error (MMSE) estimators in low SNR environments and frequency-selective fading. Another approach in \cite{OFDMCE&SD} utilizes long short-term memory (LSTM)-based DNNs to enhance channel estimation and signal detection in frequency-selective fading environments. Unlike traditional MMSE and LS methods that estimate the channel state explicitly, LSTMs enable the system to learn from sequential data, improving feature extraction and symbol detection. Simulation results indicate that LSTM-based DL achieves better or comparable performance to MMSE while offering superior adaptability in dynamically changing channel conditions. The framework used in \cite{OFDMJointCE&SD} uses a two-stage DL approach. The first stage is the channel estimation network (CENet), where the model replaces traditional interpolation techniques in pilot-based estimation schemes by treating channel estimation as an image super-resolution problem. The second stage is the channel conditioned recovery network (CCRNet), where a GAN-based architecture is used to reconstruct transmitted signals from the output of CENet. This method achieves higher accuracy compared to conventional zero-forcing (ZF), regularized zero-forcing (RZF), and MMSE methods.\\

In \cite{azari2022automated}, the authors propose replacing conventional modules in a wide-band OFDM receiver with deep neural network-based models, including MLPs and RNNs, to handle tasks like channel estimation, equalization, demapping, and decoding. This modular, end-to-end DL-based wide-band receiver outperforms traditional methods by improving BER and adapting effectively to varying channel conditions. Similarly, Pihlajasalo et al. \cite{9723316} target the challenge of demodulating OFDM signals under extreme mobility with high Doppler effects by introducing a convolutional neural network (CNN) with residual connections. This approach mitigates inter-carrier interference (ICI) and surpasses classical linear minimum mean square error (LMMSE) receivers in high-mobility scenarios. Finally, authors in \cite{9180878} present RecNet, a DL-based system that divides the OFDM receiver into two neural networks: a CNN for semi-blind channel estimation and another for signal detection based on estimated CSI. RecNet achieves superior accuracy with fewer pilots, fast convergence, and robust performance under low SNR conditions. This model offers a more efficient and adaptable solution for OFDM signal processing. Table \ref{TAB : Comparison of Deep Learning-Based OFDM Receiver Techniques} shows the summary of different Deep Leaning-based receivers used in OFDM.

\begin{table}[!h]
\caption{Comparison of Deep Learning-Based OFDM Receiver Techniques}
\label{TAB : Comparison of Deep Learning-Based OFDM Receiver Techniques}
\begin{center}
\begin{tabular}{|p{.8cm}|p{2.5cm}|p{4cm}|}
\hline
\textbf{Paper} & \textbf{Technique} & \textbf{Unique Contribution} \\ 
\hline
\cite{azari2022automated} & Modular, end-to-end DL-based wideband receiver with MLPs and RNNs & Outperforms traditional approaches, showing significant improvements in BER; demonstrates the potential of DL in capturing non-linearities and adapting to varying channel conditions. \\ 
\hline
\cite{9723316} & CNN with residual connections for demodulation under extreme mobility conditions & Significantly better performance than traditional LMMSE receivers in high mobility scenarios with severe Doppler-induced ICI. \\ 
\hline
\cite{8052521} & DNNs for end-to-end channel estimation and signal detection & Effectively handles channel distortions in OFDM systems; offers performance comparable to MMSE estimators, especially in low SNR conditions. \\ 
\hline
\cite{9180878} & Two separate neural networks: CNN for semi-blind channel estimation and another network for signal recovery & Superior accuracy in channel estimation, especially with a small number of pilots; better performance under low SNR conditions compared to traditional schemes. \\ 
\hline

\cite{OFDMCE&SD} & LSTM-Based Neural Networks for channel estimation & Improved performance over MMSE and LS, Effective for sequential data.\\
\hline
\cite{OFDMJointCE&SD} & Image-processing techniques for channel estimation, GAN-based signal recovery &  High accuracy in channel estimation and signal detection outperforming conventional ZF, RZF and MMSE.\\
\hline
\end{tabular}
\end{center}
\end{table}

\subsection{Deep Learning in MIMO Receiver}
\begin{figure}[h]
    \centering
    \includegraphics[width=1 \linewidth]{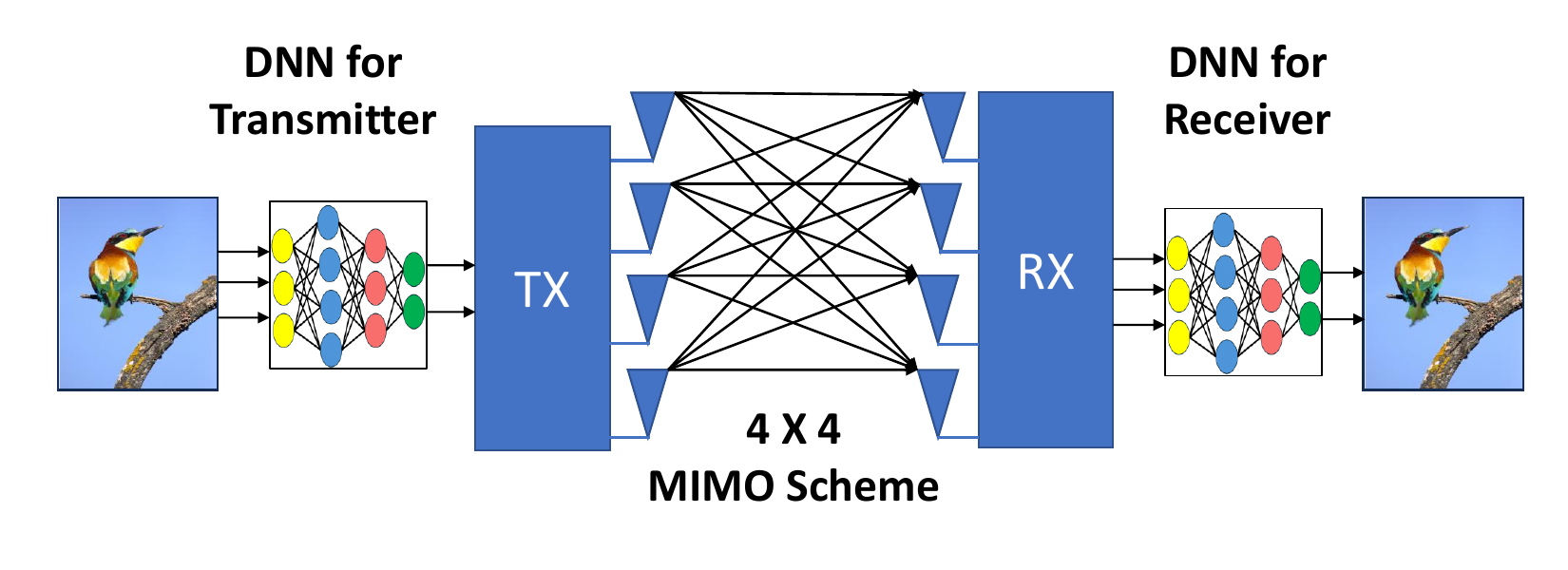}
    \caption{Schematic Diagram of Deep Learning Based MIMO system}
    \label{fig:Schematic Diagram of Deep Learning Based MIMO system}
\end{figure}

Multiple input multiple output (MIMO) systems are designed with multiple antennas at both the transmitter and receiver side to improve data throughput and reliability. MIMO systems enhance the performance of wireless communication, providing diversity gain \cite{MIMO_Diversity} and array gain \cite{MIMO_Array}. DL-based receivers offer better calibration, interference cancellation, and robustness to channel uncertainties. A genereic schematic diagram of DL-based MIMO system is shown in Figure \ref{fig:Schematic Diagram of Deep Learning Based MIMO system}.

Recent advancements in DL have addressed various challenges in MIMO wireless communication systems, improving reliability and performance under complex conditions. The study in \cite{9733027} designs an intelligent receiver based on DenseNet and MobileNet V2, replacing traditional modules with DNNs to perform end-to-end information recovery. This approach improves reliability and reduces BER across different antenna configurations, effectively managing cumulative errors. Addressing the issue of poorly calibrated decisions in DNN-based wireless receivers, Raviv et al. \cite{raviv2023modular} introduce a Bayesian learning framework combined with hybrid model-based and data-driven architectures. This work achieves superior BER and calibration compared to conventional methods by applying Bayesian principles to modular receiver designs, such as the Bayesian \textbf{DeepSIC} receiver. Similarly, authors in \cite{9242305} tackle symbol detection in multiuser MIMO systems with DeepSIC, a data-driven approach integrating DNNs into the iterative soft interference cancellation algorithm. This method eliminates the reliance on precise CSI or linear channel assumptions and enables robust performance in non-linear channels and CSI uncertainty. A summary of different deep learning approaches implemented in the receiver design of MIMO systems is presented in Table \ref{TAB: Comparison of Deep Learning-Based MIMO Receiver Techniques}.

\begin{table}[htbp]
\caption{Comparison of Deep Learning-Based MIMO Receiver Techniques}
\label{TAB: Comparison of Deep Learning-Based MIMO Receiver Techniques}
\begin{center}
\begin{tabular}{|p{.6 cm}|p{2.2cm}|p{4cm}|}
\hline
\textbf{Paper} & \textbf{Technique} & \textbf{Unique Contribution} \\ 
\hline
\cite{9733027} & Intelligent receiver based on DenseNet and MobileNet V2, replacing all receiver modules with a neural network & Higher reliability and lower BER under various conditions and antenna configurations compared to traditional receivers; improved system reliability and effective handling of cumulative errors. \\ 
\hline
\cite{raviv2023modular} & Bayesian learning with hybrid model-based/data-driven architectures for improved calibration and accuracy & Introduces modular model-based Bayesian learning for better-calibrated modules and improved overall receiver accuracy; significant improvements in BER and calibration performance over conventional methods. \\ 
\hline
\cite{9242305} & DeepSIC: data-driven approach to implement iterative soft interference cancellation using DNNs &  Outperforms model-based approaches, especially with CSI uncertainty and non-linear channels; learns joint detection without specific channel models or CSI. \\ 
\hline
\end{tabular}
\end{center}
\end{table}

\subsection{Deep Learning in Next-G Wireless Receiver}

Next-generation (Next-G) wireless communication systems face increasing challenges due to complex propagation environments and high mobility. Next-G wireless communication demands high reliability with low latency, enhanced mobile broadband, and massive integration of IoT devices. DL-based receivers provide enhanced adaptability and performance in challenging conditions such as high Doppler shifts and interference, which are prevalent in high mobility scenarios. This is a common issue for future communication, especially when IoT devices will be available everywhere.

A fully convolutional neural network, \textbf{DeepRx}, is proposed in \cite{9345504} to process wireless signals and improve performance under high Doppler shifts. It demonstrates superior performance in 5G-compliant scenarios and offers robustness in sparse pilot configurations. DeepRx performs the entire receiver pipeline from frequency domain signal stream to uncoded bits, thus replacing traditional channel estimation, equalization, and soft demapping. It provides accurate channel estimation and improved detection accuracy by using known constellation points and local symbol distribution. 

 The challenge of self-interference in full-duplex systems, which is crucial for enhancing spectral efficiency in 6G and beyond networks, is addressed in the paper \cite{shammaa2023deep}. The proposed full-duplex deep receiver (FDDR) replaces the entire information recovery process of a conventional full-duplex receiver, including self-interference cancellation, channel estimation, equalization, and demodulation with a three-hidden-layer DNN. The FDDR achieves similar BER performance as conventional receivers using Kalman Filter for channel estimation but with 80 percent less complexity. It demonstrates superior performance in scenarios with Doppler frequency mismatch, self-interference reflections, and cyclic prefix-free scenarios. The FDDR is also robust against synchronization errors and adaptable to changes in the communication environment, showing enhanced learning and signal recovery capabilities without prior channel state information.

\begin{table*}[htbp]
\caption{Summary of Deep Learning Receivers in Wireless Communication}
\label{table:deep_learning_receivers}
\centering
\footnotesize 
\begin{tabularx}{\textwidth}{|p{2.5cm}|p{1.75cm}|X|} 
\hline
\textbf{Category}&\textbf{Papers}&\textbf{Challenges Addressed}\\
\hline
Joint Optimization of Transmitter and Receiver & \cite{AuoencoderTransciever,TransciverAutoencoder2017,TranscieverAutoencoder2016,AutoencodetTransciverCNN}  & Handling different levels of mapping schemes under dynamic channel conditions, replacing traditional modulation, equalization, and demodulation methods, estimating the channel for training the transmitter and receiver jointly.  \\
\hline

Semantic Communication & \cite{qin2022semantic,beyond,Xie_2021,DeepSC-ST,DeepWive,wang2022wirelessdeepvideosemantic,semanticvideoconference} & Establishing semantic theory, dealing with poor channel conditions for semantic text communication, improving bandwidth efficiency for video transmission.\\
\hline

Task-Oriented Communication & \cite{SecureSemantic,DeeplearingdrivenTOC,tocrobotics,beyond}& Extracting relevant task information while reducing complexity, differentiating between semantic communication and task-oriented communication, ensuring timeliness and accuracy in task execution.\\
\hline

OFDM & \cite{8052521,OFDMCE&SD,OFDMJointCE&SD,azari2022automated,9723316,9180878} & Estimating CSI in missing pilots and nonlinear distortions scenario, mitigating pilot-based estimation errors and interpolation limitations, overcoming inter-carrier interference in high Doppler shift, dealing with dynamic fading environments.\\
\hline
MIMO & \cite{9733027,raviv2023modular,9242305} & Improving interference cancellation and signal detection in nonlinear channels and under CSI uncertainty, establishing end-to-end information recovery in MIMO, overcoming poorly calibrated decisions in DL-based MIMO receivers.\\
\hline
Next-G Wireless Communication & \cite{9345504,shammaa2023deep} & Improving performance in 5G by handling Doppler shifts and pilot sparsity, reducing self-interference while cutting computational complexity.\\
\hline
SIC & \cite{9837450,9348107} & Improving performance under CSI uncertainty without requiring explicit channel model knowledge, reducing computational overhead for complex channel estimation and interference cancellation.\\
\hline
SoftIC & \cite{904647,10471626} & Improving the reception and decoding of signals in environments with significant interference, mitigating the computational complexity and performance limitations of traditional model-based SoftSIC used for equalization.\\

\hline

\hline
\end{tabularx}
\end{table*}

\subsection{DL SIC Rx}

 Successive interference cancellation (SIC) is a interference mitigation technique used in communication systems. SIC works by sequentially decoding and removing signals from a composite signal, starting with the strongest signal and then progressing to the next one. This method is particularly effective in situations where the signal strengths vary significantly. DL models can enhance SIC by improving accuracy and adaptability under uncertain channel conditions.

 The application of DL to improve SIC in non-orthogonal communication systems without prior channel model knowledge is explored in the paper \cite{9837450}. The paper addresses the challenges of accurate interference cancellation in non-orthogonal downlink transmissions, particularly under conditions of uncertain CSI and rapid channel fluctuations. The proposed model, \textbf{SICNet}, employs DNNs to replace traditional hard-decision interference cancellation blocks in the SIC process, enabling soft decision-making and improved adaptability to channel variations. The introduction of SICNet offers robust performance against CSI uncertainty and does not require explicit channel model knowledge. It is also capable of adapting to changes in the number of users and their power allocations, demonstrating close performance to traditional SIC under perfect CSI conditions and superior performance under realistic CSI uncertainties.

The issues of error propagation and high computational complexity in SIC for uplink MIMO non-orthogonal multiple access (MIMO-NOMA) systems are addressed in \cite{9348107}. A DNN is used to directly decode each user's signal at every SIC step. This replaces traditional SIC components like channel estimation, signal detection, and interference cancellation. Fully connected DNNs are employed, each tailored to decode a specific user's signal in successive order without explicitly estimating channel coefficients or canceling out decoded signals. Simulation results demonstrate that the proposed method outperforms traditional SIC schemes and other DL-based approaches in terms of BER and computational efficiency.

\subsection{DL SoftIC Rx}

Soft Interference Cancellation (SoftIC) is an advanced signal processing technique used in communication systems to improve the reception and decoding of signals in environments with significant interference. Unlike methods that make hard decisions about signal values early in the processing chain, SoftIC employs a more nuanced approach, using soft values or probabilities \cite{904647}. SoftIC treats interference as noise initially and uses probabilities or soft values for interference cancellation, refining the process iteratively. This approach can yield better performance in densely packed signal environments by making use of the probabilistic information of the interference. DL-based SoftIC enhances the equalization process and reduces computational complexity.

The challenge of effective equalization in digital communication systems, specifically focusing on mitigating the computational complexity and performance limitations of traditional model-based soft interference cancellation methods, is tackled in the paper \cite{10471626}. The deep neural network-based approach, \textbf{SICNN}, replaces traditional iterative SIC methods used for equalization. This replacement addresses the inefficiencies and computational demands of conventional model-based techniques. The paper introduces SICNNv1 and SICNNv2, deep neural networks designed by deep unfolding of model-based iterative SoftIC methods. These networks aim to streamline the equalization process by integrating learned parameters that are optimized through training datasets. SICNN demonstrates superior BER performance compared to traditional and some neural network-based methods, particularly in single carrier frequency domain equalization (SC-FDE) systems. The approach significantly reduces the computational complexity and leverages novel dataset generation techniques to enhance performance at high SNR condition).\\

A summary of the challenges addressed by leveraging DL-based wireless receivers across various wireless technologies is provided in Table \ref{table:deep_learning_receivers}.


\section{Challenges and Limitations of Deep Learning in Wireless Communication} 
\label{sec:Challenes}

In Sections \ref{sec:Deep Neural Networks Architectures} and \ref{Sec: Application of Deep Learning based receivers}, we discussed the potential benefits and advantages of integrating DL into wireless receiver design. However, the practical implementation of DL presents several significant challenges. In this section, we will examine the hurdles associated with deep learning implementation in wireless communication, including issues such as data scarcity and variability,  security and privacy concerns, computational and resource constraints, integration with legacy systems, and the need for interpretability and solid theoretical foundations.

\subsection{Data Scarcity and Variability}

One of the key challenges of the DL-based wireless receiver design approach is that if large labeled datasets are unavailable, the model becomes overfitted \cite{8666641}. In wireless communication, obtaining such datasets is difficult due to data collection's high costs and logistical complexities \cite{DATA_CHallenge}. This unavailability of such datasets can lead to overfitting. An overfitted model performs well on training data but fails to generalize in real-world environments. Furthermore, differences between training data distributions and operational scenarios result in performance degradation in dynamic environments.

\subsection{Security and Privacy Concern}

Security and privacy concerns present substantial challenges when deploying DL in wireless communication systems. Data-driven approaches are more vulnerable to adversarial attacks since these attacks can manipulate input data to deceive DL models \cite{GAN_SPOOFING}. Ensuring the robustness of DL algorithms against such attacks becomes more critical when the applications involve sensitive information or critical infrastructure. Additionally, the collection and processing of user data raises privacy concerns for DL implementation in wireless systems, which makes it necessary to develop secure and privacy-preserving techniques \cite{AI_Security}.

\subsection{Computational and Resource constraints}

The computational demands of deep learning models can strain the resources of wireless devices. In edge computing scenarios, the computational power devices have limited processing power and battery life \cite{RIS-Assisted}. To deal with the power constraint challenge, it becomes necessary to design lightweight models that can operate efficiently in resource-constrained environments while still delivering high performance. The challenge of balancing model complexity with operational efficiency is a critical area of ongoing research.

\subsection{integration with legacy systems}
Most wireless systems today are designed around modular and deterministic algorithms. Implementing DL-based wireless receivers into the existing framework raises the issue of compatibility. One potential solution is to transition from traditional methods to DL-based approaches by developing hybrid frameworks that integrate both paradigms. These hybrid frameworks can serve as an intermediate stage as DL-based wireless communication undergoes research and implementation to address the various challenges associated with fully deploying DL-based wireless frameworks.

\subsection{Interpretability and Theoretical Foundations}
Traditional wireless communication methods are developed based on solid mathematical foundations. When performance anomalies arise, it becomes easier to analyze the issues using a solid theoretical foundation. In contrast,  the "black-box" nature of DL models limits their interpretability. The interpretability concern of DL models makes it challenging to analyze when there is a sudden degradation in performance \cite{DL_Paradigm}. As a result, the acceptance of DL-based approaches is hindered in applications that demand transparency and accountability.

\section{Conclusion} 
\label{sec:conclusion}

This survey highlights the diverse applications of deep learning-based receivers. This study shows that DL-based wireless receivers outperform traditional model-based methods in addressing critical challenges such as channel estimation, interference cancellation, and signal demodulation. Moreover, our study demonstrates the efficacy of DL-based receivers in various modern wireless technologies, such as semantic communication, task-oriented communication, OFDM, MIMO, Next-G networks, SIC, and SoftIC. 

The integration of advanced architectures such as CNNs, RNNs, GANs, and autoencoders has enabled unprecedented efficiency and reliability by facilitating end-to-end learning and optimization of receiver functionalities. Despite these advancements, several challenges remain, including the need for large labeled datasets, security and privacy concerns, computational and resource constraints, integration with legacy technologies, and the interpretability of DL models. Addressing these challenges requires further research into data-efficient training techniques, model explainability, and hardware-friendly implementations to ensure the widespread deployment of DL-based receivers in future wireless systems.

As DL-based wireless receivers are drawing attention among researchers, DL architectures are set to play a crucial role in shaping the receivers of future wireless networks. The ongoing research in this area will pave the way for more intelligent, adaptive, and high-performance communication systems. Our survey paper will serve as a valuable resource for researchers in this field.


\begin{thebibliography}{100}

\bibitem{DeepReciverdataAug}
T.~Raviv and N.~Shlezinger, ``{Adaptive data augmentation for deep receivers},'' {\em 2022 IEEE 23rd International Workshop on Signal Processing Advances in Wireless Communication (SPAWC)}, 2022.

\bibitem{DeepReciverdatahybrid}
T.~Raviv, S.~Park, O.~Simeone, Y.~C. Eldar, and N.~Shlezinger, ``{Online meta-learning for hybrid model-based deep receivers},'' {\em IEEE Transactions on Wireless Communications}, vol.~22, pp.~6415--6431, 2023.

\bibitem{DL_Overview}
J.~Schmidhuber, ``{Deep learning in neural networks: An overview},'' {\em Neural Networks}, vol.~61, p.~85–117, Jan. 2015.

\bibitem{DLforWireless}
T.~Erpek, T.~O'Shea, Y.~Sagduyu, Y.~Shi, and T.~Clancy, {\em {Deep Learning for Wireless Communications}}, pp.~223--266.
\newblock 01 2020.

\bibitem{8666641}
C.~Zhang, P.~Patras, and H.~Haddadi, ``{Deep Learning in Mobile and Wireless Networking: A Survey},'' {\em IEEE Communications Surveys \& Tutorials}, vol.~21, no.~3, pp.~2224--2287, 2019.

\bibitem{Aldossari2019MachineLF}
S.~A. Aldossari and K.-C. Chen, ``{Machine Learning for Wireless Communication Channel Modeling: An Overview},'' {\em Wireless Personal Communications}, vol.~106, pp.~41 -- 70, 2019.

\bibitem{9665363}
J.~Jiao, X.~Sun, L.~Fang, and J.~Lyu, ``{An overview of wireless communication technology using deep learning},'' {\em China Communications}, vol.~18, no.~12, pp.~1--36, 2021.

\bibitem{SyncRcvRCV}
J.~Peng, L.~Zhang, and D.~McLernon, ``{Ms-assisted receiver-receiver time synchronization strategy for femtocells},'' {\em 2011 IEEE 73rd Vehicular Technology Conference (VTC Spring)}, 2011.

\bibitem{SyncFreq}
A.~Markus, F.~Wermke, F.~Winkler, and B.~Meffert, ``{Frequency synchronization for wireless networks using field programmable gate arrays},'' {\em 2016 IEEE International Conference on Electronics, Circuits and Systems (ICECS)}, 2016.

\bibitem{SyncWSN}
P.~Wang and S.~W. Xu, ``{Research on time synchronization algorithm in wireless sensor networks},'' {\em Advanced Materials Research}, vol.~403-408, pp.~1397--1400, 2011.

\bibitem{SyncNatural}
M.~Harashima, Y.~Hiroyuki, and M.~Hasegawa, ``{Synchronization of wireless sensor networks using natural environmental signals based on noise-induced phase synchronization phenomenon},'' {\em 2012 IEEE 75th Vehicular Technology Conference (VTC Spring)}, 2012.

\bibitem{Syncdistributed}
M.~J. Kim, S.~J. Maeng, and Y.~S. Cho, ``{Distributed Synchronization Technique for OFDMA-Based Wireless Mesh Networks Using a Bio-Inspired Algorithm},'' {\em Sensors}, 2015.

\bibitem{SyncGPS}
D.~Pallier, V.~L. Cam, and S.~Pillement, ``{Energy-Efficient GPS Synchronization for Wireless Nodes", journal = "IEEE Sensors Journal},'' 2021.

\bibitem{chestimationsurvey}
A.~Q. Jumaah~Althahab and S.~A. Kadhim~Alrufaiaat, ``{A Comprehensive Review on Various Estimation Techniques for Multi Input Multi Output Channel},'' {\em Journal of University of Babylon for Engineering Sciences}, 2019.

\bibitem{chestimationFIR}
C.~Yu and G.~Gui, ``{Recursive Least Square–based Fast Sparse Multipath Channel Estimation},'' {\em International Journal of Communication Systems}, 2017.

\bibitem{ChannelEstimatonLMMSE}
Z.~Tong, M.~Guo, X.~Yang, and W.~H. Zhang, ``{Performance Comparison of LS and LMMSE Channel Estimation Algorithm for CO-OFDM System},'' {\em Applied Mechanics and Materials}, 2011.

\bibitem{CHestimationML5G}
H.~A. Le, T.~V. Chien, T.~H. Nguyen, H.~Choo, and V.~D. Nguyen, ``{Machine Learning-Based 5g-and-Beyond Channel Estimation for MIMO-OFDM Communication Systems},'' {\em Sensors}, 2021.

\bibitem{chestimationCompressive}
Z.~He, L.~Zhou, Y.~Yang, Y.~Chen, X.~Ling, and C.~Liu, ``{Compressive Sensing-Based Channel Estimation for FBMC-OQAM System Under Doubly Selective Channels},'' {\em IEEE Access}, 2019.

\bibitem{ChestimationINM}
Z.~Yang, Y.~Inoue, J.~Wan, and L.~Chen, ``{Channel Parameters Identification Based on IMM Algorithm for Variant Correlation Channel},'' {\em Mathematical Problems in Engineering}, 2015.

\bibitem{ChestimationOrthoMIMO}
L.~Wang, ``{Channel Estimation and Combining Orthogonal Pilot Design in MIMO-OFDM System},'' {\em Journal of Networks}, 2014.

\bibitem{ChestimationDL}
Y.~Hao, G.~Y. Li, and B.~Juang, ``{Power of Deep Learning for Channel Estimation and Signal Detection in OFDM Systems},'' {\em IEEE Wireless Communications Letters}, 2018.

\bibitem{equalizationAdaptiveKovac}
L.~Kovács, J.~Levendovszky, A.~Oláh, and G.~Treplan, ``{Approximate minimum bit error rate equalization for fading channels},'' {\em EURASIP Journal on Advances in Signal Processing}, vol.~2010, 2010.

\bibitem{equalizationAdaptiveXi}
Z.~Xi, ``{Analysis of adaptive equalization algorithms},'' {\em Highlights in Science, Engineering and Technology}, vol.~70, pp.~295--305, 2023.

\bibitem{equalizationDBN}
X.~Liu, J.~Zhang, S.~Gao, W.~Tong, Y.~Wang, M.~Lei, B.~Hua, Y.~Cai, Y.~Zou, and M.~Zhu, ``{Demonstration of 144-gbps photonics-assisted thz wireless transmission at 500 ghz enabled by joint dbn equalizer},'' {\em Micromachines}, vol.~13, p.~1617, 2022.

\bibitem{EqualizationMMSE-DFE}
W.~Hua, Y.~Huang, J.~Du, and L.~Li, ``{A universal mmse-dfe equalizer with its application to wlan receiver},'' {\em Wireless Personal Communications}, vol.~85, pp.~2507--2518, 2015.

\bibitem{EqualizationLinear}
A.~Kumar, A.~Tiwari, and R.~S. Mishra, ``{Linear block equalizers in rayleigh fading channel with normalized channel impulse response},'' {\em International Journal of Computer Applications}, vol.~93, pp.~21--26, 2014.

\bibitem{equalizationParityCheck}
Q.~Yu, L.~Tang, H.~Lin, and F.~Cao, ``{Parity-check coding transmit diversity for wireless communications with high mobility},'' {\em IEEE Transactions on Vehicular Technology}, vol.~71, pp.~1737--1749, 2022.

\bibitem{SpacetimecodingMimo}
H.~Li, L.~W. Ang, S.~Palaniappan, and J.~Wang, ``{Enhancing MIMO Capacity Through Space-Time Coding: Analysis And Design Framework},'' {\em Journal of Informatics and Web Engineering}, vol.~2, pp.~49--55, 2023.

\bibitem{SpaceTimecodingMetasurface}
T.~J. Cui, S.~Liu, G.~D. Bai, and Q.~Ma, ``{Direct transmission of digital message via programmable coding metasurface},'' {\em Research}, vol.~2019, 2019.

\bibitem{SpaceTimecodingEvolutio}
N.~A. Kaimkhani, Z.~Chen, and F.~Yin, ``{Evolution of diversity gain with and without coding gain in mimo for emerging wireless networks},'' {\em International Journal of Computer Theory and Engineering}, vol.~9, pp.~32--37, 2017.

\bibitem{DemodulationCNN}
J.~Wang, H.~Huang, J.~Liu, and J.~Li, ``{Joint demodulation and error correcting codes recognition using convolutional neural network},'' {\em IEEE Access}, vol.~10, pp.~104844--104851, 2022.

\bibitem{DemodulationDL2019}
H.~Wang, Z.~Wu, S.~Ma, S.~Lu, H.~Zhang, G.~Ding, and S.~Li, ``{Deep learning for signal demodulation in physical layer wireless communications: prototype platform, open dataset, and analytics},'' {\em IEEE Access}, vol.~7, pp.~30792--30801, 2019.

\bibitem{DemodulationDL22}
A.~Ahmad, S.~Agarwal, S.~Darshi, and S.~Chakravarty, ``{Deepdemod: bpsk demodulation using deep learning over software-defined radio},'' {\em IEEE Access}, vol.~10, pp.~115833--115848, 2022.

\bibitem{ChannelcodingDL2019}
R.~Fritschek, R.~F. Schaefer, and G.~Wunder, ``{Deep learning for channel coding via neural mutual information estimation},'' {\em 2019 IEEE 20th International Workshop on Signal Processing Advances in Wireless Communications (SPAWC)}, 2019.

\bibitem{Channelcoding2012}
M.~K. Arti, R.~K. Mallik, and R.~Schober, ``{Joint channel estimation and decoding of space-time block codes in af mimo relay networks},'' {\em 2012 International Conference on Signal Processing and Communications (SPCOM)}, pp.~1--5, 2012.

\bibitem{ChannelcodingDLMimo17}
T.~J. O’Shea, T.~Erpek, and T.~C. Clancy, ``{Deep learning based mimo communications},'' 2017.

\bibitem{Sourcecoding2020}
N.~Rajapaksha and N.~Rajatheva, ``{Low complexity autoencoder based end-to-end learning of coded communications systems},'' {\em 2020 IEEE 91st Vehicular Technology Conference (VTC2020-Spring)}, 2020.

\bibitem{SourcecodingAutoencoders2021}
N.~A. Letizia and A.~M. Tonello, ``{Capacity-driven autoencoders for communications},'' {\em IEEE Open Journal of the Communications Society}, vol.~2, pp.~1366--1378, 2021.

\bibitem{Zheng_2021}
S.~Zheng, S.~Chen, and X.~Yang, ``{DeepReceiver: A Deep Learning-Based Intelligent Receiver for Wireless Communications in the Physical Layer},'' {\em IEEE Transactions on Cognitive Communications and Networking}, vol.~7, p.~5–20, Mar. 2021.

\bibitem{Deep_Learning}
Y.~LeCun, Y.~Bengio, and G.~Hinton, ``{Deep Learning},'' {\em Nature}, vol.~521, pp.~436--44, 05 2015.

\bibitem{MLPwsn}
G.~Reddy, S.~Kadiyala, C.~Potluri, S.~Saravanan, G.~Athisha, M.~M~a, and M.~Sujaritha, ``{An Intrusion Detection Using Machine Learning Algorithm Multi-Layer Perceptron (MlP): A Classification Enhancement in Wireless Sensor Network (WSN)},'' {\em International Journal on Recent and Innovation Trends in Computing and Communication}, vol.~10, pp.~139--145, 12 2022.

\bibitem{MLPmmw}
Y.~Zheng, B.~Huang, and Z.~Lu, ``{MLP-mmWP: High-Precision Millimeter Wave Positioning Based on MLP-Mixer Neural Networks},'' {\em Sensors}, vol.~23, no.~8, 2023.

\bibitem{CE_MLP}
M.~Andari, ``{Pilot Based Channel Estimation in Broadband Power Line Communication Networks},'' {\em Communications and Network}, vol.~04, pp.~240--247, 01 2012.

\bibitem{Distortion_MLP}
L.~A.~M. Pereira, L.~L. Mendes, C.~J.~A. Bastos-Filho, and S.~Arismar~Cerqueira, ``{Linearization Schemes for Radio Over Fiber Systems Based on Machine Learning Algorithms},'' {\em IEEE Photonics Technology Letters}, vol.~34, no.~5, pp.~279--282, 2022.

\bibitem{AMC_MLP}
H.~C. Dubey, Nandita, and A.~K. Tiwari, ``{Blind modulation classification based on MLP and PNN},'' in {\em 2012 Students Conference on Engineering and Systems}, pp.~1--6, 2012.

\bibitem{ResourceAllocation_MLP}
R.~B. Ch and N.~Sendrayan, ``{An adaptive MLP-based joint optimization of resource allocation and relay selection in device-to-device communication using hybrid meta-heuristic algorithm},'' {\em EURASIP Journal on Wireless Communications and Networking}, vol.~2024, 06 2024.

\bibitem{article1}
D.~M~N, M.~R. Kounte, and C.~Ravindra, ``{Design of a Deep Learning based Intelligent Receiver for a Wireless Communication System},'' {\em International Journal of Electrical and Electronics Research}, vol.~12, pp.~228--237, 03 2024.

\bibitem{CNNCE}
Y.~Jin, J.~Zhang, S.~Jin, and B.~Ai, ``{Channel Estimation for Cell-Free mmWave Massive MIMO Through Deep Learning},'' {\em IEEE Transactions on Vehicular Technology}, vol.~68, pp.~10325--10329, Oct 2019.

\bibitem{CNNCEBeamspace}
H.~He, C.-K. Wen, S.~Jin, and G.~Y. Li, ``{Deep Learning-Based Channel Estimation for Beamspace mmWave Massive MIMO Systems},'' {\em IEEE Wireless Communications Letters}, vol.~7, no.~5, pp.~852--855, 2018.

\bibitem{cnnchannelestimationofdm}
A.~S. M.~M. Jameel, A.~Malhotra, A.~E. Gamal, and S.~Hamidi-Rad, ``{Deep OFDM Channel Estimation: Capturing Frequency Recurrence},'' 2024.

\bibitem{CNNSpecSensing}
E.~Vijay and K.~Aparna, {\em {CNN Depending Spectrum Sensing for Effective Data Transmission in Wireless Communication}}, pp.~35--46.
\newblock 10 2023.

\bibitem{CNNinterference}
H.~Zhang, M.~Zhao, M.~Zhang, S.~Lin, Y.~Dong, and H.~Wang, ``{A combination network of CNN and transformer for interference identification},'' {\em Frontiers in Computational Neuroscience}, vol.~17, 12 2023.

\bibitem{CE_CNN_MASSIVEMIMO}
P.~Dong, H.~Zhang, G.~Y. Li, I.~Gaspar, and N.~Naderializadeh, ``{Deep CNN-Based Channel Estimation for mmWave Massive MIMO Systems},'' {\em IEEE Journal of Selected Topics in Signal Processing}, vol.~13, pp.~989--1000, 2019.

\bibitem{CE_CNN_RIS}
N.~K. Kundu and M.~R. McKay, ``{Channel Estimation for Reconfigurable Intelligent Surface Aided MISO Communications: From LMMSE to Deep Learning Solutions},'' {\em IEEE Open Journal of the Communications Society}, vol.~2, pp.~471--487, 2021.

\bibitem{Demod_CNN}
M.~A.~S. Sejan, M.~H. Rahman, and H.~Song, ``{Demod-CNN: A Robust Deep Learning Approach for Intelligent Reflecting Surface-Assisted Multiuser MIMO Communication},'' {\em Sensors}, vol.~22, p.~5971, 2022.

\bibitem{Decod_Vid_CNN}
V.~Boussard, S.~Coulombe, F.-X. Coudoux, P.~Corlay, and A.~Trioux, ``{CRC-Based Multi-Error Correction of H.265 Encoded Videos in Wireless Communications},'' in {\em 2021 International Conference on Visual Communications and Image Processing (VCIP)}, pp.~1--5, 2021.

\bibitem{Decod_PolarCodes_CNN}
T.~Chen, A.~K. Ho, C.~D. Wu, S.~Wong, and A.~Wu, ``{Convolutional Neural Network-Aided Bit-Flipping for Belief Propagation Decoding of Polar Codes},'' {\em ICASSP 2021 - 2021 IEEE International Conference on Acoustics, Speech and Signal Processing (ICASSP)}, 2021.

\bibitem{MobileNet}
A.~G. Howard, M.~Zhu, B.~Chen, D.~Kalenichenko, W.~Wang, T.~Weyand, M.~Andreetto, and H.~Adam, ``{MobileNets: Efficient Convolutional Neural Networks for Mobile Vision Applications},'' 2017.

\bibitem{MobiNet_2023}
S.~Hua, Q.~Wang, and X.~Xu, ``Application of machine learning in wireless communication,'' {\em Theoretical and Natural Science}, vol.~12, pp.~130--135, 2023.

\bibitem{Response_MOBI}
H.~Fan, D.~W.~L. Tay, and A.~Ang, ``{Intelligent Pick-and-Place System Using MobileNet},'' {\em Electronics}, vol.~12, p.~621, 2023.

\bibitem{RNN5GCE}
I.~Taha, ``{Channel state information estimation for 5G wireless communication systems: recurrent neural networks approach},'' {\em PeerJ Computer Science}, vol.~7, 08 2021.

\bibitem{RNNCoChanIn}
M.~A. Sejan, M.~Rahman, M.~A. Aziz, R.~Tabassum, Y.-H. You, D.~Hwang, and H.-K. Song, ``{Interference Management for a Wireless Communication Network Using a Recurrent Neural Network Approach},'' {\em Mathematics}, vol.~12, pp.~1--17, 06 2024.

\bibitem{RNNOFDMIM}
M.~A. Aziz, M.~Rahman, M.~A. Sejan, R.~Tabassum, D.~Hwang, and H.-K. Song, ``{Deep Recurrent Neural Network Based Detector for OFDM with Index Modulation},'' {\em IEEE Access}, vol.~PP, 06 2024.

\bibitem{RNNMassive}
J.~P. Lemayian and J.~M. Hamamreh, ``{Massive {MIMO} {Channel} {Prediction} {Using} {Recurrent} {Neural} {Networks}},'' {\em RS Open Journal on Innovative Communication Technologies}, vol.~1, aug 12 2020.
\newblock https://rs-ojict.pubpub.org/pub/massive-mimo-channel-prediction-using-recurrent-neural-networks.

\bibitem{CE_RNN}
M.~H.~E. Ali, F.~Alraddady, M.~Y. Al-Thunaibat, and S.~Elnazer, ``{Machine learning-based channel state estimators for 5g wireless communication systems},'' {\em Computer Modeling in Engineering \& Amp; Sciences}, vol.~135, pp.~755--778, 2023.

\bibitem{CH_DEC_RNN_2022}
J.~Kim, S.~Hosseinalipour, T.~Kim, D.~J. Love, and C.~G. Brinton, ``{Linear coding for gaussian two-way channels},'' {\em 2022 58th Annual Allerton Conference on Communication, Control, and Computing (Allerton)}, 2022.

\bibitem{CH_DEC_RNN_kim2018}
H.~Kim, Y.~Jiang, R.~Rana, S.~Kannan, S.~Oh, and P.~Viswanath, ``{Communication Algorithms via Deep Learning},'' 2018.

\bibitem{CH_DEC_RNN_2018}
R.~Sattiraju, A.~Weinand, and H.~D. Schotten, ``{Performance analysis of deep learning based on recurrent neural networks for channel coding},'' {\em 2018 IEEE International Conference on Advanced Networks and Telecommunications Systems (ANTS)}, pp.~1--6, 2018.

\bibitem{AMC_RNN}
 .~Kaladè, L.~H. Crockett, and R.~W. Stewart, ``{Using Sequence to Sequence Learning for Digital BPSK and QPSK Demodulation},'' {\em 2018 IEEE 5G World Forum (5GWF)}, pp.~317--320, 2018.

\bibitem{ISI_RNN}
M.~Mosavat and G.~Montorsi, ``{Single-frequency network terrestrial broadcasting with 5gnr numerology using recurrent neural network},'' {\em Electronics}, vol.~11, p.~3130, 2022.

\bibitem{Demod_RNN_ASK}
Y.~Xu, S.~Yang, H.~Li, J.~Lv, and N.~Bai, ``{Adaptive noise-resistant low-power ask demodulator design in uhf rfid chips},'' {\em Electronics}, vol.~10, p.~3168, 2021.

\bibitem{generativeadversarialnetworks}
I.~J. Goodfellow, J.~Pouget-Abadie, M.~Mirza, B.~Xu, D.~Warde-Farley, S.~Ozair, A.~Courville, and Y.~Bengio, ``{Generative Adversarial Networks},'' 2014.

\bibitem{GANsprectrumsensing}
K.~Davaslioglu and Y.~E. Sagduyu, ``{Generative Adversarial Learning for Spectrum Sensing},'' in {\em 2018 IEEE International Conference on Communications (ICC)}, pp.~1--6, 2018.

\bibitem{GANAMC}
M.~Patel, X.~Wang, and S.~Mao, ``{Data augmentation with conditional GAN for automatic modulation classification},'' in {\em Proceedings of the 2nd ACM Workshop on Wireless Security and Machine Learning}, WiseML '20, (New York, NY, USA), p.~31–36, Association for Computing Machinery, 2020.

\bibitem{GANRadioanamoly}
X.~Zhou, J.~Xiong, X.~Zhang, X.~Liu, and J.~Wei, ``{A Radio Anomaly Detection Algorithm Based on Modified Generative Adversarial Network},'' {\em IEEE Wireless Communications Letters}, vol.~10, no.~7, pp.~1552--1556, 2021.

\bibitem{GANDLEndtoEnd}
Y.~Hao, L.~Liang, G.~Y. Li, and B.~Juang, ``{Deep learning-based end-to-end wireless communication systems with conditional gans as unknown channels},'' {\em IEEE Transactions on Wireless Communications}, vol.~19, pp.~3133--3143, 2020.

\bibitem{Autoencodesurvey}
P.~Li, Y.~Pei, and J.~Li, ``{A comprehensive survey on design and application of autoencoder in deep learning},'' {\em Applied Soft Computing}, vol.~138, p.~110176, 2023.

\bibitem{AutoencodrPhysicalLayer}
T.~J. O’Shea, T.~Roy, N.~West, and B.~C. Hilburn, ``{Physical layer communications system design over-the-air using adversarial networks},'' {\em 2018 26th European Signal Processing Conference (EUSIPCO)}, 2018.

\bibitem{AuoencoderTransciever}
S.~Dörner, S.~Cammerer, J.~Hoydis, and S.~t. Brink, ``{Deep Learning Based Communication Over the Air},'' {\em IEEE Journal of Selected Topics in Signal Processing}, vol.~12, no.~1, pp.~132--143, 2018.

\bibitem{AutoendoerJsccImage}
E.~Bourtsoulatze, D.~B. Kurka, and D.~Gündüz, ``{Deep joint source-channel coding for wireless image transmission},'' {\em ICASSP 2019 - 2019 IEEE International Conference on Acoustics, Speech and Signal Processing (ICASSP)}, pp.~4774--4778, 2019.

\bibitem{AutoencoderJsccBER}
N.~O. Chikezie, U.~C. Femi, O.~O. Ozioma, A.~E. Oluwatomisin, A.~Chukwuebuka, N.~E. Onyekachi, and G.~C. Kalejaiye, ``{Ber performance evaluation using deep learning algorithm for joint source channel coding in wireless networks},'' {\em Advances in Science, Technology and Engineering Systems Journal}, vol.~7, pp.~127--139, 2022.

\bibitem{TransciverAutoencoder2017}
T.~O’Shea and J.~Hoydis, ``{An Introduction to Deep Learning for the Physical Layer},'' {\em IEEE Transactions on Cognitive Communications and Networking}, vol.~3, no.~4, pp.~563--575, 2017.

\bibitem{TranscieverAutoencoder2016}
T.~J. O'Shea, K.~Karra, and T.~C. Clancy, ``{Learning to communicate: Channel auto-encoders, domain specific regularizers, and attention},'' in {\em 2016 IEEE International Symposium on Signal Processing and Information Technology (ISSPIT)}, pp.~223--228, 2016.

\bibitem{AutoencodetTransciverCNN}
B.~Zhu, J.~Wang, L.~He, and J.~Song, ``{Joint Transceiver Optimization for Wireless Communication PHY Using Neural Network},'' {\em IEEE Journal on Selected Areas in Communications}, vol.~37, no.~6, pp.~1364--1373, 2019.

\bibitem{ChannelAgnosticEnd-to-End}
H.~Ye, G.~Y. Li, B.-H.~F. Juang, and K.~Sivanesan, ``{Channel Agnostic End-to-End Learning Based Communication Systems with Conditional GAN},'' in {\em 2018 IEEE Globecom Workshops (GC Wkshps)}, pp.~1--5, 2018.

\bibitem{approximatingvoidlearningstochastic}
T.~J. O'Shea, T.~Roy, and N.~West, ``{Approximating the Void: Learning Stochastic Channel Models from Observation with Variational Generative Adversarial Networks},'' 2018.

\bibitem{WGAN-BASEDtraining}
S.~Dörner, M.~Henninger, S.~Cammerer, and S.~ten Brink, ``{WGAN-based Autoencoder Training Over-the-air},'' in {\em 2020 IEEE 21st International Workshop on Signal Processing Advances in Wireless Communications (SPAWC)}, pp.~1--5, 2020.

\bibitem{s24102993}
C.~P. Davey, I.~Shakeel, R.~C. Deo, and S.~Salcedo-Sanz, ``{Deep Learning Based Over-the-Air Training of Wireless Communication Systems without Feedback},'' {\em Sensors}, vol.~24, no.~10, 2024.

\bibitem{qin2022semantic}
Z.~Qin, X.~Tao, J.~Lu, W.~Tong, and G.~Y. Li, ``{Semantic Communications: Principles and Challenges},'' {\em arXiv preprint arXiv:2201.01389}, 2022.

\bibitem{beyond}
D.~Gündüz, Z.~Qin, I.~Aguerri, H.~Dhillon, Z.~Yang, A.~Yener, K.~Wong, and C.-B. Chae, ``{Beyond Transmitting Bits: Context, Semantics, and Task-Oriented Communications},'' 07 2022.

\bibitem{Xie_2021}
H.~Xie, Z.~Qin, G.~Y. Li, and B.-H. Juang, ``{Deep Learning Enabled Semantic Communication Systems},'' {\em IEEE Transactions on Signal Processing}, vol.~69, p.~2663–2675, 2021.

\bibitem{DeepSC-ST}
Z.~Weng, Z.~Qin, X.~Tao, C.~Pan, G.~Liu, and G.~Li, ``{Deep Learning Enabled Semantic Communications With Speech Recognition and Synthesis},'' {\em IEEE Transactions on Wireless Communications}, vol.~PP, pp.~1--1, 09 2023.

\bibitem{DeepWive}
T.-Y. Tung and D.~Gündüz, ``{DeepWiVe: Deep-Learning-Aided Wireless Video Transmission},'' 11 2021.

\bibitem{wang2022wirelessdeepvideosemantic}
S.~Wang, J.~Dai, Z.~Liang, K.~Niu, Z.~Si, C.~Dong, X.~Qin, and P.~Zhang, ``{Wireless Deep Video Semantic Transmission},'' 2022.

\bibitem{semanticvideoconference}
P.~Jiang, C.-K. Wen, S.~Jin, and G.~Y. Li, ``{Wireless Semantic Communications for Video Conferencing},'' {\em IEEE Journal on Selected Areas in Communications}, vol.~41, no.~1, pp.~230--244, 2023.

\bibitem{SecureSemantic}
Z.~Yang, M.~Chen, G.~Li, Y.~Yang, and Z.~Zhang, ``{Secure Semantic Communications: Fundamentals and Challenges},'' {\em IEEE Network}, vol.~38, no.~6, pp.~513--520, 2024.

\bibitem{tocrobotics}
R.~Detry, J.~Papon, and L.~Matthies, ``{Task-oriented grasping with semantic and geometric scene understanding},'' in {\em 2017 IEEE/RSJ International Conference on Intelligent Robots and Systems (IROS)}, pp.~3266--3273, 2017.

\bibitem{DeeplearingdrivenTOC}
Y.~Sagduyu, S.~Ulukus, and A.~Yener, ``{Age of Information in Deep Learning-Driven Task-Oriented Communications},'' pp.~1--6, 05 2023.

\bibitem{8052521}
H.~Ye, G.~Y. Li, and B.-H. Juang, ``{Power of Deep Learning for Channel Estimation and Signal Detection in OFDM Systems},'' {\em IEEE Wireless Communications Letters}, vol.~7, no.~1, pp.~114--117, 2018.

\bibitem{OFDMCE&SD}
K.~J. Wong, F.~H. Juwono, and R.~Reine, ``{Deep Learning for Channel Estimation and Signal Detection in OFDM-Based Communication Systems},'' {\em ELKHA}, 2022.

\bibitem{OFDMJointCE&SD}
X.~Yi and C.~Zhong, ``{Deep Learning for Joint Channel Estimation and Signal Detection in OFDM Systems},'' {\em IEEE Communications Letters}, vol.~24, p.~2780–2784, Dec. 2020.

\bibitem{azari2022automated}
B.~Azari, H.~Cheng, N.~Soltani, H.~Li, Y.~Li, M.~Belgiovine, T.~Imbiriba, S.~D’Oro, T.~Melodia, Y.~Wang, {\em et~al.}, ``{Automated deep learning-based wide-band receiver},'' {\em Computer Networks}, vol.~218, p.~109367, 2022.

\bibitem{9723316}
J.~Pihlajasalo, D.~Korpi, M.~Honkala, J.~M.~J. Huttunen, T.~Riihonen, J.~Talvitie, M.~A. Uusitalo, and M.~Valkama, ``{Deep Learning Based OFDM Physical-Layer Receiver for Extreme Mobility},'' in {\em 2021 55th Asilomar Conference on Signals, Systems, and Computers}, pp.~395--399, 2021.

\bibitem{9180878}
C.~Liu and T.~Arslan, ``{RecNet: Deep Learning-Based OFDM Receiver with Semi-Blind Channel Estimation},'' in {\em 2020 IEEE International Symposium on Circuits and Systems (ISCAS)}, pp.~1--4, 2020.

\bibitem{MIMO_Diversity}
K.~Rosengren and P.-S. Kildal, ``{Radiation efficiency, correlation, diversity gain and capacity of a six-monopole antenna array for a MIMO system: theory, simulation and measurement in reverberation chamber},'' {\em IEE Proceedings-Microwaves, Antennas and Propagation}, vol.~152, no.~1, pp.~7--16, 2005.

\bibitem{MIMO_Array}
F.~Rusek, D.~Persson, B.~K. Lau, E.~G. Larsson, T.~L. Marzetta, O.~Edfors, and F.~Tufvesson, ``{Scaling Up MIMO: Opportunities and Challenges with Very Large Arrays},'' {\em IEEE Signal Processing Magazine}, vol.~30, no.~1, pp.~40--60, 2013.

\bibitem{9733027}
B.~Wang, K.~Xu, S.~Zheng, H.~Zhou, and Y.~Liu, ``{A Deep Learning-Based Intelligent Receiver for Improving the Reliability of the MIMO Wireless Communication System},'' {\em IEEE Transactions on Reliability}, vol.~71, no.~2, pp.~1104--1115, 2022.

\bibitem{raviv2023modular}
T.~Raviv, S.~Park, O.~Simeone, and N.~Shlezinger, ``{Modular model-based bayesian learning for uncertainty-aware and reliable deep MIMO receivers},'' in {\em 2023 IEEE International Conference on Communications Workshops (ICC Workshops)}, pp.~1032--1037, IEEE, 2023.

\bibitem{9242305}
N.~Shlezinger, R.~Fu, and Y.~C. Eldar, ``{DeepSIC: Deep Soft Interference Cancellation for Multiuser MIMO Detection},'' {\em IEEE Transactions on Wireless Communications}, vol.~20, no.~2, pp.~1349--1362, 2021.

\bibitem{9345504}
M.~Honkala, D.~Korpi, and J.~M.~J. Huttunen, ``{DeepRx: Fully Convolutional Deep Learning Receiver},'' {\em IEEE Transactions on Wireless Communications}, vol.~20, no.~6, pp.~3925--3940, 2021.

\bibitem{shammaa2023deep}
M.~Shammaa, M.~Mashaly, and A.~El-mahdy, ``{A deep learning-based adaptive receiver for full-duplex systems},'' {\em AEU-International Journal of Electronics and Communications}, vol.~170, p.~154822, 2023.

\bibitem{9837450}
T.~Van~Luong, N.~Shlezinger, C.~Xu, T.~M. Hoang, Y.~C. Eldar, and L.~Hanzo, ``{Deep Learning Based Successive Interference Cancellation for the Non-Orthogonal Downlink},'' {\em IEEE Transactions on Vehicular Technology}, vol.~71, no.~11, pp.~11876--11888, 2022.

\bibitem{9348107}
M.~A. Aref and S.~K. Jayaweera, ``{Deep Learning-aided Successive Interference Cancellation for MIMO-NOMA},'' in {\em GLOBECOM 2020 - 2020 IEEE Global Communications Conference}, pp.~1--5, 2020.

\bibitem{904647}
W.-J. Choi, K.-W. Cheong, and J.~Cioffi, ``{Iterative soft interference cancellation for multiple antenna systems},'' in {\em 2000 IEEE Wireless Communications and Networking Conference. Conference Record (Cat. No.00TH8540)}, vol.~1, pp.~304--309 vol.1, 2000.

\bibitem{10471626}
S.~Baumgartner, O.~Lang, and M.~Huemer, ``{SICNN: Soft Interference Cancellation Inspired Neural Network Equalizers},'' {\em IEEE Transactions on Machine Learning in Communications and Networking}, vol.~2, pp.~384--406, 2024.

\bibitem{DATA_CHallenge}
H.~Wu, X.~Li, and Y.~Deng, ``{Deep learning-driven wireless communication for edge-cloud computing: opportunities and challenges},'' {\em Journal of Cloud Computing Advances Systems and Applications}, vol.~9, 04 2020.

\bibitem{GAN_SPOOFING}
Y.~Shi, K.~Davaslioglu, and Y.~E. Sagduyu, ``{Generative Adversarial Network in the Air: Deep Adversarial Learning for Wireless Signal Spoofing},'' {\em IEEE Transactions on Cognitive Communications and Networking}, vol.~7, no.~1, pp.~294--303, 2021.

\bibitem{AI_Security}
X.~Qiu, Z.~Du, and X.~Sun, ``{Artificial Intelligence-Based Security Authentication: Applications in Wireless Multimedia Networks},'' {\em IEEE Access}, vol.~PP, pp.~1--1, 11 2019.

\bibitem{RIS-Assisted}
K.~K. Nguyen, A.~Masaracchia, V.~Sharma, H.~V. Poor, and T.~Q. Duong, ``Ris-assisted uav communications for iot with wireless power transfer using deep reinforcement learning,'' {\em IEEE Journal of Selected Topics in Signal Processing}, vol.~16, no.~5, pp.~1086--1096, 2022.

\bibitem{DL_Paradigm}
L.~Dai, R.~Jiao, F.~Adachi, H.~V. Poor, and L.~Hanzo, ``{Deep Learning for Wireless Communications: An Emerging Interdisciplinary Paradigm},'' {\em IEEE Wireless Communications}, vol.~27, no.~4, pp.~133--139, 2020.

\end{thebibliography}

\end{document}